%% file: main.tex
\UseRawInputEncoding
\documentclass{article}

\usepackage[T1]{fontenc}
\usepackage[utf8]{inputenc}

\usepackage[margin=1in]{geometry}
\usepackage{pifont}
\usepackage{amsmath}
\usepackage{amssymb}
\usepackage{array}
\usepackage{graphicx}
\usepackage{epstopdf}
\usepackage{tikz}
\usepackage{booktabs}
\usepackage{subfig}
\usepackage{siunitx}
\usepackage{caption}
\usepackage{cases}
\usepackage{bm}
\usepackage{titlesec}
\usepackage{authblk}
\usepackage[percent]{overpic}
\usepackage[numbers,round,sort&compress]{natbib}
\usepackage[colorlinks,linkcolor=black,citecolor=blue,urlcolor=blue]{hyperref} 
\let\cite\citep

\newcommand{\beginsupplement}{%
  \setcounter{table}{0}\renewcommand{\thetable}{S\arabic{table}}%
  \setcounter{figure}{0}\renewcommand{\thefigure}{S\arabic{figure}}%
  \renewcommand{\theHtable}{S\arabic{table}}%
  \renewcommand{\theHfigure}{S\arabic{figure}}%
}

\DeclareCaptionFont{mysize}{\fontsize{8}{9.6}\selectfont}
\graphicspath{{./}{./figures/}}

\newcommand{\sflabel}[1]{{\footnotesize\bf\sffamily #1}}

\title{\textbf{Hebbian self-organization of monosynaptic reflexes in a
biologically realistic multiarticular limb}}

\author[1]{Jonathan Realmuto}
\author[1,2,3,4]{Erfan Nozari}
\author[5]{Gerald E. Loeb}
\affil[1]{Department of Mechanical Engineering, University of California, Riverside, CA}
\affil[2]{Department of Electrical and Computer Engineering, University of California, Riverside, CA}
\affil[3]{Department of Bioengineering, University of California, Riverside, CA}
\affil[4]{Neuroscience Graduate Program, University of California, Riverside, CA}
\affil[5]{Alfred E. Mann Department of Biomedical Engineering, University of Southern California, Los Angeles, CA}

\date{}

\begin{document}

\maketitle
\begin{abstract}
\noindent Spinal monosynaptic reflexes depend on patterns of connectivity
between muscle spindle afferents and motoneurons that should reflect
musculoskeletal dynamics. One possibility is that the details of these circuits
are specified genetically, but this requires an unlikely coevolution of
musculoskeletal and neural elements. Here we show how biomechanical structure
can determine the connectivity of self-organized spinal circuits. Using a
neuromechanical model combining a multi-joint limb, proprioceptive feedback,
and Hebbian plasticity driven by motor babbling, we find that biologically
realistic connectivity emerges spontaneously when babbling is neither too
sparse nor too frequent. This results in strong homonymous projections, weaker
heteronymous projections among functional synergists, and very weak connections
among antagonists. This outcome also depends on musculoskeletal morphometry.
Heteronymous synergies form when proximal muscles are strongest---the
characteristic gradient of mammalian limbs---whereas excessively strong
biarticular muscles lead to unphysiological antagonist coupling. The mechanisms
differ: homonymous connections arise from direct sensorimotor correlation
during fusimotor coactivation by beta motoneurons, while heteronymous
connections depend on multi-joint dynamics that emerge only after homonymous
reflexes stabilize and increasing muscle strength generates multisegmental
motion. Overall, this study provides a mechanistic basis for fetal sensorimotor
development and generates testable predictions about the consequences of
naturally occurring and experimentally induced variations. \end{abstract}

\noindent\textbf{Significance Statement.} How does the spinal cord acquire the
precise Ia-to-motoneuron connectivity that supports short-latency reflexes
within and across muscles in adults? One view is that genes specify each
connection; another, that experience shapes the wiring. Using a computational
model of a limb, its muscles, and its sensory feedback, we show that
spontaneous motor babbling, together with a simple Hebbian learning rule, can
generate realistic reflex connectivity. The pattern that emerges depends on the
body: naturalistic connections arise when muscle strengths follow the mammalian
proximal-to-distal gradient, and change predictably when that gradient is
altered. We propose that adult connectivity reflects both genetic rules and
experiential system identification. This principle has implications for motor
development, neurological disease, and the design of embodied robots.

\subsubsection*{Introduction} \label{sec:intro}

Muscle spindle primary afferents (Ia) connect strongly to motoneurons (MNs) of
their own muscle, more weakly to functional synergists, and rarely to
antagonists~\cite{eccles1957synaptic, fritz1989pattern, meunier1993pattern,
  marchand2000monosynaptic, mcclelland2001short, hultborn2006spinal,
yoshimoto2022monosynaptic}. This pattern underpins the stretch reflex and
coordinates multi-joint responses to
perturbation~\cite{sherrington1906integrative, nichols1994biomechanical}. Many
genetic mechanisms control sequences and types of neurogenesis and interneuronal
connectivity~\cite{ladle2007assembly}, but the source of such muscle-specific
connectivity remains contentious. Genetic specification of such details poses a
long-known problem for musculoskeletal
evolution: any musculoskeletal mutation can survive only if another mutation occurs simultaneously and congruently in the neuronal circuitry that controls the affected musculoskeletal plant~\cite{baldwin1896new,baldwin1897organic}. Self-organization during
spontaneous fetal motor activity has been proposed as an alternative and
demonstrated in a highly simplified model system~\cite{enander2022model1}. Such
refinement is thought to begin before birth: spontaneous fetal movements
(twitches) pair motor output with sensory feedback, generating correlated
sensorimotor activity implicated in shaping developing
circuits~\cite{blankenship2010mechanisms, petersson2003spontaneous}. This
activity continues into early postnatal life as sleep twitches and spontaneous,
open-ended movements~\cite{blumberg2015developing, kanazawa2023open}.

A natural candidate mechanism for this refinement is Hebbian plasticity, where
synapses are strengthened when pre- and postsynaptic activity are correlated,
and weakened when such correlations are absent or
inconsistent~\cite{hebb1949organization, bi1998synaptic}. In the developing
sensorimotor system, these correlations are shaped by the body itself. Muscle
mechanics, limb geometry, and spindle sensitivity constrain which Ia neurons
and MNs are co-active during spontaneous movement. Fusimotor control regulates
spindle sensitivity and can thereby shape the statistics available for Hebbian
refinement~\cite{friese2009gamma}. As a result, fetal and early postnatal
movements may provide structured sensorimotor statistics that reinforce
homonymous and synergistic Ia--MN pathways while leaving antagonistic pathways
weak---the normal adult pattern. In this view, the musculoskeletal system does
not merely execute spinal commands; its mechanics result in the neural activity
patterns from which spinal connectivity can be learned.

Hebbian learning has been successfully modeled to account for emergent
functionality in many parts of the central nervous
system~\cite{kuniyoshi2006early,der2015novel} using the Oja rule
algorithm~\cite{oja1982simplified}. The Hebbian principle of pre- and
postsynaptic coactivity faces a basic obstacle to account for the homonymous
Ia--MN synapse. When a motoneuron activates its muscle fibers, the contractile
force shortens and unloads the spindles that are anchored in parallel with
them, silencing the homonymous Ia afferents while the MN is
active~\cite{hunt1951stretch}. This problem has been overcome by proposing and
modeling an ``anti-Oja''
rule~\cite{marques2014spontaneous,petersson2003spontaneous} but there is no
physiological support for such a plasticity mechanism in mammals. Another such
model claimed to use a positive Oja rule but then inverted the sign of the MN
activity, which effectively creates a similarly unphysiological negative
learning rule~\cite{almanzor2025self}.

A more complete model of known motor innervation can overcome the above
problem. Mature mammalian muscles are innervated by both $\alpha$MNs, which
innervate only extrafusal muscle fibers, and $\gamma$MNs, which innervate only
intrafusal muscle fibers whose contractions stretch and activate Ia receptors.
The $\gamma$MNs are probably not functional during early fetal development, but
a persistent fraction of the extrafusal innervation is by skeletofusimotor
$\beta$MNs that also activate intrafusal muscle fibers. During fetal
development, all MNs in one motor nucleus are electrotonically linked by gap
junctions~\cite{personius2007reduced} and generate spontaneous activity that is
thus synchronized across them. This is responsible for the large, random muscle
twitches that occur throughout fetal and early postnatal development. While
extrafusal fibers are immature and too weak to move the limb so as to shorten
and unload the spindle afferents, the fusimotor component of any $\beta$MN
activity would likely produce a net increase in spindle firing rather than
silencing~\cite{enander2022model1}.

\begin{figure*}[t]
  \centering
  \begin{overpic}[width=\linewidth]{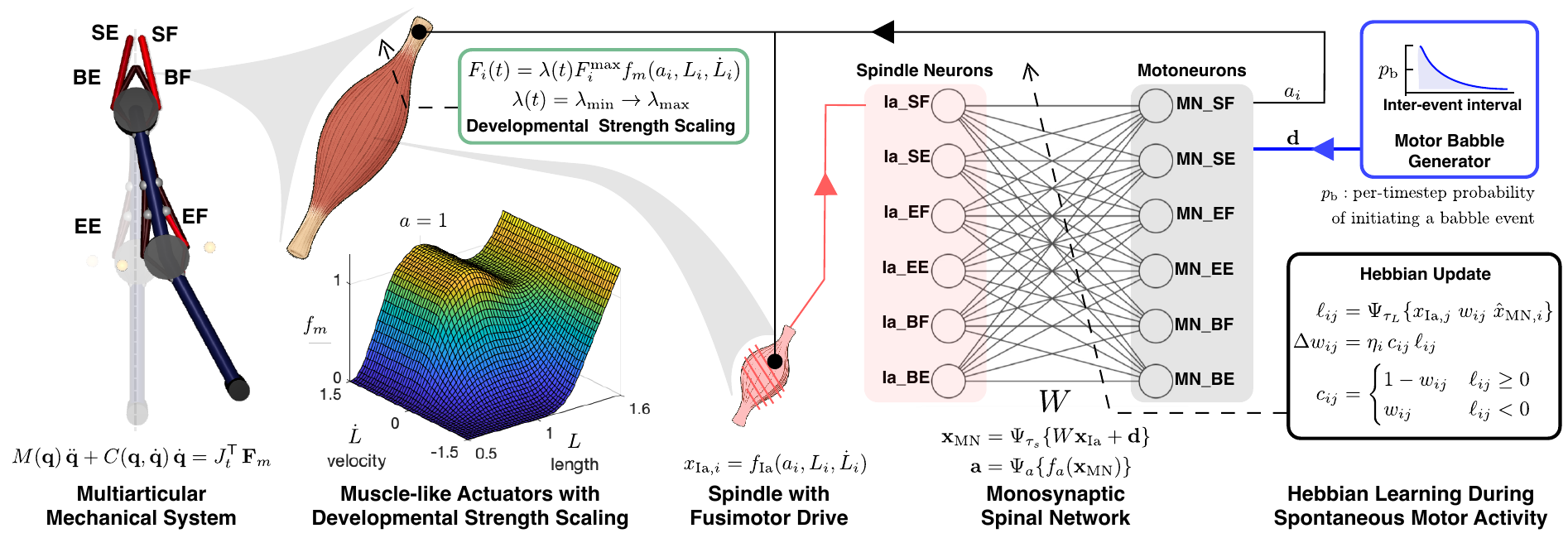}
    {\footnotesize
    \put(0.5,1.5){\sflabel{A}}
    \put(17.75,1.5){\sflabel{B}}
    \put(42,1.5){\sflabel{C}}
    \put(60.5,1.5){\sflabel{D}}
    \put(79,1.5){\sflabel{E}}}
  \end{overpic}
\caption{\textbf{A neuromechanical model of spinal self-organization.} A
  closed sensorimotor loop where spontaneous motor babble drives a
  musculoskeletal limb with Hebbian plasticity self-organizing the monosynaptic
  spinal network. (\sflabel{A}) Multiarticular mechanical system: a two-joint
  limb with six muscles---four monoarticular (shoulder/elbow flexors and
  extensors SF, SE, EF, EE) and two biarticular (BF, BE)---under rigid-body
  dynamics. (\sflabel{B}) Muscle-like actuators with developmental strength
  scaling: muscle force follows a force--length--velocity relation $f_{\mathrm
  m}$ (surface shown at activation $a=1$), scaled by a strength gain
  $\lambda(t)$ that matures from $\lambda_{\min}$ to $\lambda_{\max}$.
  (\sflabel{C}) Muscle spindles under fusimotor drive produce Ia afferent
  signals encoding muscle activation, length, and velocity. (\sflabel{D})
  Monosynaptic spinal network: Ia afferents (spindle neurons) project to
  motoneurons through the plastic weight matrix $W$; the motoneuron cell
  potential $x_{\mathrm{MN}}$ integrates reflex input with descending babble
  drive $\mathbf{d}$ and sets muscle activation $\mathbf{a}$. (\sflabel{E})
  Hebbian learning during spontaneous motor activity: the babble generator
  emits drive $\mathbf{d}$ as a renewal process (onset probability $p_{\mathrm
  b}$ per time step, giving approximately exponential waiting times between
  events); weights $w_{ij}$ update by a soft-bounded, eligibility-filtered
  Hebbian rule (eligibility $\ell_{ij}$, soft bound $c_{ij}$, learning rate
  $\eta_i$). \emph{Notation:} $f_{(\cdot)}$ are static (instantaneous) maps;
  $\Psi_{\tau_{(\cdot)}}$ are linear first-order causal filters with time
  constants $\tau_{\mathrm L}$ (learning), $\tau_{\mathrm H}$ (phasic
  high-pass), and $\tau_{\mathrm s}$ (membrane); $\Psi_{\mathrm a}$ is the
  muscle-activation filter (first-order nonlinear dynamics). In \sflabel{E},
  $\hat x_{\mathrm{MN},i}$ abbreviates the postsynaptic term of
  Eq.~\eqref{eq:hebbian}, the phasic motoneuron activity scaled by the Hebbian
gain minus its tonic mean, $\kappa\,\Psi_{\tau_{\mathrm
H}}(x_{\mathrm{MN},i})-\bar x_{\mathrm{MN},i}$.} \label{fig:model}
\end{figure*}

\medskip \noindent\textbf{A neuromechanical model for spinal
self-organization.} Enander and colleagues~\cite{enander2022model1,
enander2022model2} showed that Hebbian learning with a positive Oja rule during
simulated motor babbling produces canonical spinal motifs: homonymous stretch
reflexes and mature types of spinal interneurons that emerge without genetic
pre-specification. Their model was extremely simple, however, with only
monoarticular limbs operated by simple antagonist muscle pairs. The present
study extends the Hebbian plasticity model to a musculoskeletal system with two
segments linked by six muscles. We test the hypothesis that the complete
physiological pattern of Ia--MN activity in homonymous, synergist and antagonist
muscles can be produced by Hebbian plasticity as the spontaneous random
twitches become stronger and more coordinated by the emerging reflex circuits.
The model, summarized in Fig.~\ref{fig:model}, consists of a planar arm with
two joints and six muscles: monoarticular shoulder flexor (SF), shoulder
extensor (SE), elbow flexor (EF), elbow extensor (EE), and two biarticular
muscles spanning both joints (BF, BE). It operates in a viscous environment
simulating the water-filled, muscular womb. For each muscle, a spindle model
receives muscle length, muscle velocity, and fusimotor drive, producing a Ia
afferent signal. In our model fusimotor effects are captured by making the Ia
signal depend on activation of its homonymous MN as well as length and velocity
of its parent muscle ($x_{\mathrm{Ia}}=f_{\mathrm{Ia}}(a,L,\dot L)$). Each Ia
afferent projects to all MNs through a plastic weight matrix $W$ that is
initially randomized and subsequently modified as a result of the Hebbian
learning rule. MN activity is the sum of the Ia feedback plus a spontaneous
drive signal $\mathbf{d}$, modeled as
\begin{equation} \label{eq:mn}
\mathbf{x}_\mathrm{MN}
=
\Psi_{\tau_{\mathrm s}} \{ W\mathbf{x}_\mathrm{Ia} + \mathbf{d} \},
\end{equation}
where $\Psi_{\tau_{\mathrm s}}$ is a low-pass filter with time constant
$\tau_{\mathrm s}$ representing membrane integration. Each row of $W$ defines
the proprioceptive input field of one MN and MN activity is converted to a
bounded neural command by a clipped affine transformation, introducing an
activation threshold and saturating output range. This command is passed to the
MuJoCo simulator's muscle actuator, which applies physiological activation
dynamics and force--length--velocity properties before generating muscle
force~\cite{todorov2012mujoco}. The resulting force moves the limb, changing
muscle kinematics and their effects on the spindle models, closing the
sensorimotor loop.

During simulation, force capacity develops gradually through a developmental
scaling factor $\lambda(t)$, as it would during fetal development. Strength is
held constant within each 90~s episode and increased by equal steps between
episodes, rising from $\lambda_\mathrm{min}$ to $\lambda_\mathrm{max}$ over the
first half of the simulation and held at $\lambda_\mathrm{max}$ thereafter.
Thus, $\lambda_\mathrm{max}$ sets the endpoint of strength development.

Embodiment is defined by both the distribution of muscle strength and the
distribution of limb mass. Muscle strength is parameterized by the relative
maximum isometric force across muscle groups,
\[
\mathrm{SR} =
F^\mathrm{max}_\mathrm{S}:F^\mathrm{max}_\mathrm{E}:F^\mathrm{max}_\mathrm{B},
\]
where S, E, and B denote shoulder, elbow, and biarticular muscles. Limb inertia
is parameterized by the proximal-to-distal mass ratio,
\[
\mathrm{MR} = m_\mathrm{proximal}:m_\mathrm{distal}.
\]
With total strength and total mass held fixed, changing $\mathrm{SR}$ alters
the force-generating consequences of spontaneous activity, whereas changing
$\mathrm{MR}$ alters the inertial consequences of the same activity.

Spontaneous activity is generated by motor babbling. Each motoneuron
independently receives brief excitatory pulses (50--100~ms, random amplitude)
whose onset occurs with probability $p_{\mathrm b}$ per simulation step, so
$p_{\mathrm b}$ sets the frequency of spontaneous twitches.

Synaptic weights $W$ evolve via a modified rule based on Enander et
al.~\cite{enander2022model1}. The eligibility signal ($\ell_{ij}$) integrates
the Ia--MN correlation over time with weights updated as:
\begin{align} 
  \ell_{ij} &=
  \Psi_{\tau_{\mathrm L}} \bigl( x_{\mathrm{Ia},j} \, w_{ij} \,
  (\kappa \, \Psi_{\tau_{\mathrm H}}(x_{\mathrm{MN},i}) -
\bar{x}_{\mathrm{MN},i})\bigr) \label{eq:hebbian} \\
\Delta w_{ij} &= \eta_i \, c_{ij} \, \ell_{ij} \nonumber \\
c_{ij} &= \begin{cases} 1 - w_{ij} & \ell_{ij}\geq 0 \nonumber \\ w_{ij} &
\ell_{ij}<0 \end{cases} \\
\eta_i &= \bar{x}_{\mathrm{MN},i}^4 \; \eta \nonumber
\end{align}
where $\Psi_{\tau_{\mathrm H}}(\cdot)$ is a high-pass filter extracting phasic
transients in motoneuron activity $x_{\mathrm{MN},i}$,
$\bar{x}_{\mathrm{MN},i}$ is a slow moving average of baseline activity,
$\kappa$ is the \emph{Hebbian gain}, $\Psi_{\tau_{\mathrm L}}(\cdot)$ is the
learning filter used to integrate the pre- and postsynaptic correlations, and
$\eta_i$ is the learning rate. The $\bar{x}_{\mathrm{MN},i}^4$ term gates
plasticity by the tonic activity of the postsynaptic motoneuron, suppressing
updates to its inputs when it is inactive, and $c_{ij}$ soft-bounds weights to
$[0,1]$. When $\kappa=1$, the postsynaptic term matches that of the Enander
rule, and an input is potentiated only when its coactivity with phasic
motoneuron transients exceeds the tonic baseline. For $\kappa>1$, phasic
transients are weighted more strongly relative to that baseline, lowering the
threshold for potentiation and favoring inputs coactive with dynamic motoneuron
activity over those that are merely tonically coactive, which are depressed.

\begin{figure*}[!t]
  \centering
  \begin{overpic}[width=\linewidth]{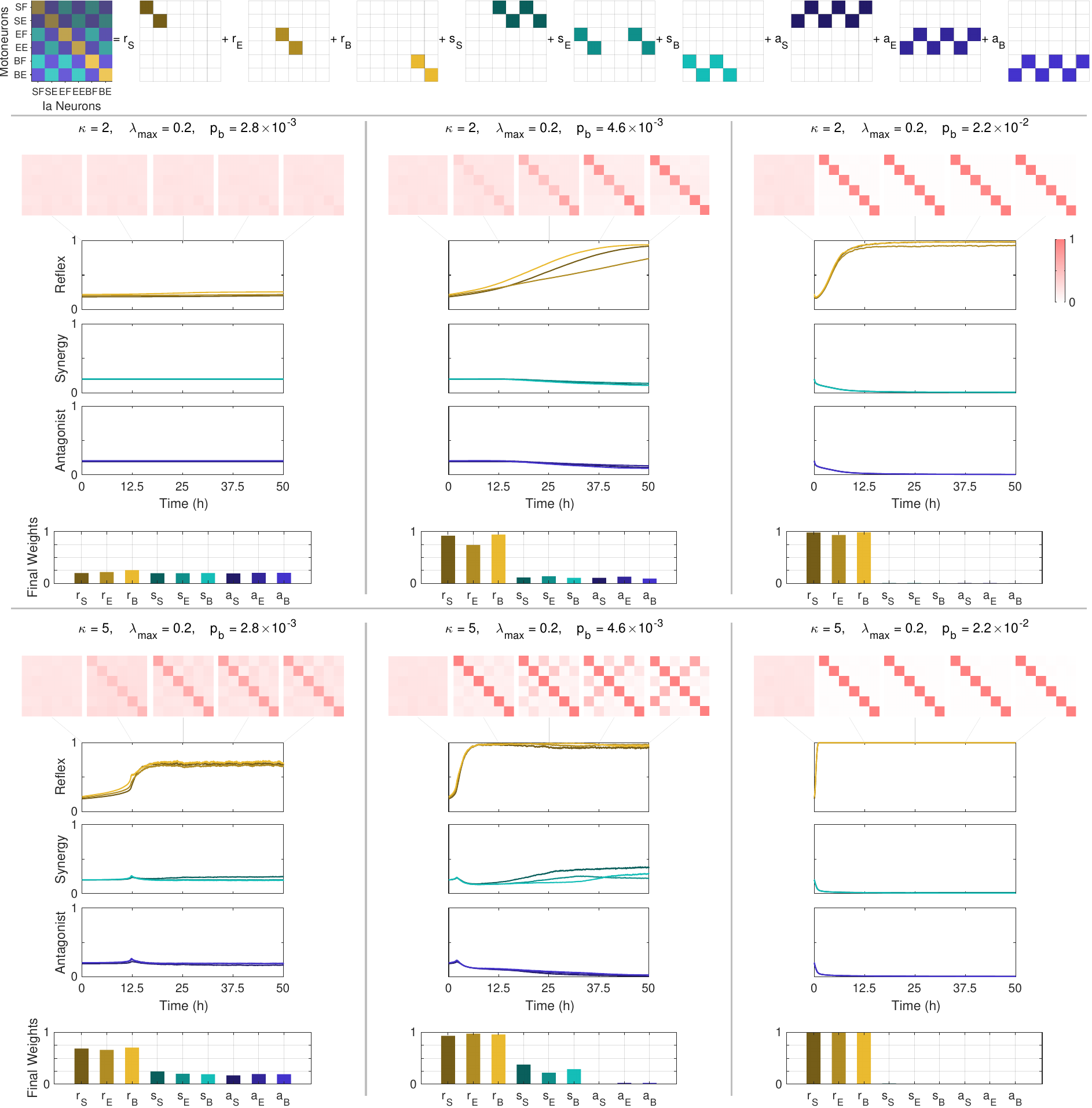}
    \put(1,90.5){\sflabel{A}}
    \put(3,87){\sflabel{B}}
    \put(36,87){\sflabel{C}}
    \put(69,87){\sflabel{D}}
    \put(3,43){\sflabel{E}}
    \put(36,43){\sflabel{F}}
    \put(69,43){\sflabel{G}}
  \end{overpic}
    \caption{%
\textbf{Self-organized Ia$\to$MN connectivity produces human-like spinal motifs
within an intermediate babble regime.} (\sflabel{A}) Each self-organized
connectivity matrix $W$ is approximated by regression as a weighted sum of
muscle-specific submotifs associated with the shoulder (S), elbow (E), and
biarticular (B) muscles: homonymous reflex motifs ($r_{\mathrm S},r_{\mathrm
E},r_{\mathrm B}$), heteronymous synergist motifs ($s_{\mathrm S},s_{\mathrm
E},s_{\mathrm B}$), and antagonist motifs ($a_{\mathrm S},a_{\mathrm
E},a_{\mathrm B}$). The resulting \emph{motif spectrum} summarizes the relative
expression of each circuit motif. (\sflabel{B}--\sflabel{G}) Example learning
outcomes: each column is a babble probability $p_{\mathrm b}$ (low,
intermediate, high) and each row a Hebbian gain $\kappa$, at fixed global
strength ceiling $\lambda_{\max}$; heat-map sequences show the learned $W$
matrix over learning, time courses the motif coefficients, and bar plots their
final values. (\sflabel{B}--\sflabel{D}) At $\kappa=2$ the circuit does not
organize at low babble (\sflabel{B}) and develops only reflex-biased
connectivity, with synergist and antagonist weights undistinguished, at
intermediate and high babble (\sflabel{C}, \sflabel{D}).
(\sflabel{E}--\sflabel{G}) At $\kappa=5$, self-organization becomes
babble-dependent: low babble (\sflabel{E}) yields incomplete homonymous
reflexes; intermediate babble (\sflabel{F}) the human-like combination of
reflex and heteronymous synergy ($R>S>A$); high babble (\sflabel{G}) reflexes
alone, with synergy pruned.} \label{fig:motif-regression}
\end{figure*}

\subsubsection*{Results}

\smallskip \noindent\textbf{A motif decomposition reveals human-like spinal
organization resulting from Hebbian plasticity.}\label{sec:result1} To
characterize the connectivity that emerges from babble-driven plasticity, we
approximate the self-organized $6\times6$ weight matrix $W$ through linear
regression, as a weighted sum of a set of connectivity \emph{motifs}
(Fig.~\ref{fig:motif-regression}\sflabel{A}): a \emph{homonymous reflex} motif,
in which each Ia afferent excites its own motoneuron; a heteronymous
\emph{synergist} motif, coupling functionally related muscles; and an
\emph{antagonist} motif, coupling muscles with opposing action. Each motif is
resolved per muscle class---shoulder (S), elbow (E), and biarticular
(B)---yielding nine coefficients $(r_\mathrm{S},r_\mathrm{E},r_\mathrm{B},
s_\mathrm{S},s_\mathrm{E},s_\mathrm{B},
a_\mathrm{S},a_\mathrm{E},a_\mathrm{B})$ whose values form a compact
\emph{motif spectrum}. Averaging across pools collapses these to three scalars,
overall reflex ($R$), synergy ($S$), and antagonist ($A$) strength.

Realistic organization requires a Hebbian gain $\kappa>1$. With $\kappa=1$ (the
original Enander rule), weights in our multi-joint limb changed little over the
simulated developmental period, showing at most a weak reflex bias
(Figs.~\ref{fig:si-kappa-lines},~\ref{fig:si-kappa-heatmaps}). At $\kappa=2$,
nothing organizes at low babble probability, and at intermediate and high
babble homonymous reflexes form but synergist and antagonist connections are
not distinguished (Fig.~\ref{fig:motif-regression}\sflabel{B}--\sflabel{D}). We
therefore use $\kappa=5$ throughout
(Fig.~\ref{fig:motif-regression}\sflabel{E}--\sflabel{G}). At low babble
probability (Fig.~\ref{fig:motif-regression}\sflabel{E}), reflexes form only
partially and synergist and antagonist weights remain near their initial
values: activity is too sparse to complete organization within the 50~h
developmental window that we simulated.

At intermediate babble probability, self-organization produces a strongly
ordered spectrum, $R > S > A \approx 0$
(Fig.~\ref{fig:motif-regression}\sflabel{F}). The motifs emerge in sequence.
Homonymous reflexes saturate within the first few hours, while muscle strength
is still a small fraction of its ceiling: movement-driven length and velocity
signals are weak relative to each afferent's own activation-dependent
(fusimotor) component, so each afferent correlates mainly with its own
motoneuron. Heteronymous synergies then grow slowly over tens of hours, and
antagonist connections are gradually pruned. This emergent ordering reproduces
the general features of heteronymous Ia connectivity as mapped
electrophysiologically in the human leg~\cite{meunier1993pattern} and the cat
forelimb~\cite{fritz1989pattern}: homonymous excitation is strongest,
heteronymous excitation between synergists is present but weaker, and
projections onto antagonists are weak or absent.

At high babble probability (Fig.~\ref{fig:motif-regression}\sflabel{G}),
synergist connections are pruned along with antagonists, leaving homonymous
reflexes alone. The two motif classes thus differ sharply in robustness:
reflexes form across more than an order of magnitude of $p_{\mathrm b}$,
whereas synergies appear only within a roughly threefold window. Together with
their late emergence, this suggests that heteronymous synergies, unlike
homonymous reflexes, depend on limb movement produced by sufficiently strong
muscles, a mechanistic distinction that we examine next.


\begin{figure*}[t!]
  \centering
  \includegraphics[width=\linewidth]{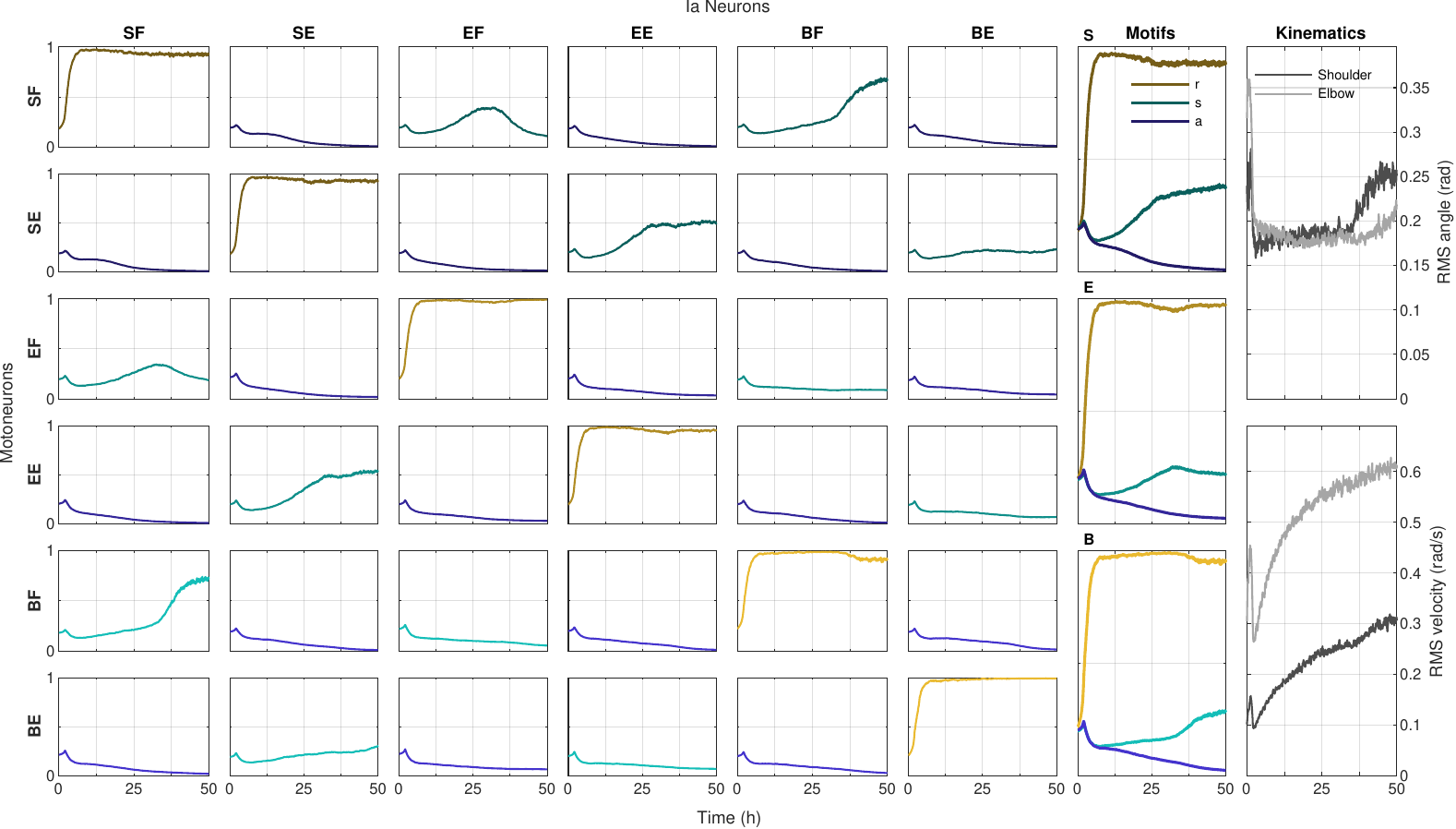}
 \caption{\textbf{Development of individual Ia$\to$MN connections.} Weight
   $w_{ij}(t)$ of every connection for an example run ($\kappa=5$,
   $\lambda_{\max}=0.2$, $p_{\mathrm b}=4.6\times10^{-3}$); rows index
   motoneurons, columns Ia afferents. Traces are colored by motif class and
   muscle pool as in Fig.~\ref{fig:motif-regression}\sflabel{A}. Homonymous
   reflexes saturate within the first hours. Synergist connections diverge:
   some grow late in development, some rise and then decline, and some remain
   near their initial value. Antagonist connections decay throughout.
   \emph{Motifs}: pool motif coefficients (reflex $r$, synergy $s$, antagonist
   $a$) for the shoulder (S), elbow (E), and biarticular (B) pools.
   \emph{Kinematics}: RMS joint angle (excursion from the rest posture) and RMS
 joint velocity per 90~s episode, smoothed over 11 episodes. Joint velocity
falls as reflexes form, then rises as muscle strength develops; joint excursion
changes comparatively little.} \label{fig:gc-grid}
\end{figure*}

\begin{figure*}[t]
  \centering
  \begin{overpic}[width=0.95\linewidth]{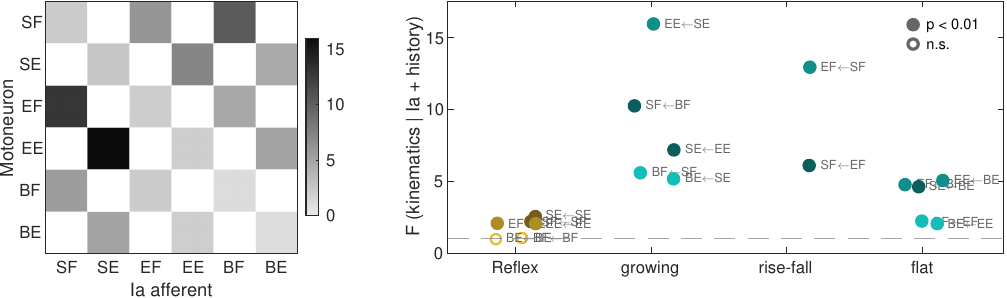}
    {\footnotesize
    \put(-2,28){\sflabel{A}}
    \put(42,28){\sflabel{B}}}
  \end{overpic}
  \caption{%
\textbf{Kinematic contribution to weight growth.} For each connection, we
performed a Granger causality test of whether 0.4~s of joint position and
velocity history improves prediction of weight-increase events beyond the
presynaptic Ia afferent and the history of the weight change itself. Events
are local peaks of the positive weight change while the weight lies below 90\%
of its maximum. Rows index motoneurons, columns Ia afferents; antagonist
connections, which all decay, were not tested (blank). (\sflabel{A}) $F$
statistic. (\sflabel{B}) $F$ by connection class, one dot per connection;
synergist connections are grouped by their trajectory in Fig.~\ref{fig:gc-grid}
(growing, rising then declining, or remaining near their initial value). Filled
dots, $p<0.01$ after Bonferroni correction for 18 tests; dashed line, $F=1$,
the expectation with no kinematic contribution.
}\label{fig:gc-partialR2}
\end{figure*}

\smallskip
\noindent\textbf{Homonymous and heteronymous connections form by different
organizing principles.}\label{sec:mechanism}

The late emergence and fragility of synergies suggest that the two motif
classes are built from different correlations: homonymous reflexes from a
muscle's own activation-coupled afferent signal, available while muscles are
still weak, and heteronymous synergies from multi-joint limb movement that only
sufficiently strong muscles can produce. We examined this first by following
individual connections through development, and then by asking whether limb
\emph{kinematics} predict weight growth beyond the presynaptic afferent.

Individual connections develop on distinct timescales (Fig.~\ref{fig:gc-grid};
raw signals in Fig.~\ref{fig:si-timehistory} and Movies~S1--S5). Homonymous
reflexes saturate within the first hours, while muscle strength is still low;
joint velocity falls as they form. Synergist connections change slowly and
later, as strength develops and joint velocity rises, and antagonist
connections decay throughout. Joint excursion changes comparatively little, so
the later period is distinguished mainly by faster movements. Synergist
connections are also heterogeneous: some grow steadily, some rise and then
decline, and some remain near their initial value. Homonymous reflex
connectivity may decline slightly but is generally preserved even after the
muscles become strong enough to move the limb (Fig.~\ref{fig:gc-grid}). When
this multiarticular system is twitching frequently, a twitch in one muscle may
tend to occur when the limb is swinging so as to lengthen or shorten the muscle
and its spindles. The spindle unloading effects that always occur in a
monoarticular system such as studied by Enander et al.~\cite{enander2022model1}
are washed out by the twitches that happen to occur when a muscle is stretched
by the inertial limb motion and the fusimotor effect then augments Ia activity.

To test whether movement drives this later growth, we asked, for each
connection, whether recent joint position and velocity improve prediction of
weight-increase events beyond the presynaptic Ia afferent and the history of
the weight change (Granger causality test; Fig.~\ref{fig:gc-partialR2}).
Because the afferent already encodes its own muscle's length and velocity, any
added contribution reflects movement not captured locally, such as mechanically
coupled motion of the other joint. We restricted the test to growth events,
occurring before a weight reaches 90\% of its maximum. Antagonist connections,
which all decay, were not tested. Kinematics added little predictive value for
homonymous connections ($F=1.0$--$2.6$; 4 of 6 reached $p<0.01$,
Bonferroni-corrected) and more for synergist connections that formed
($F=5.2$--$16.0$; all $p<0.01$), most strongly from shoulder afferents to elbow
motoneurons. Homonymous reflexes are thus learned from correlations local to
each muscle, whereas synergist connections are learned from the movement of the
whole limb.

Together, these analyses point to two organizing principles of our model
system. Homonymous reflexes arise early from activation-coupled (fusimotor)
correlations that depend little on multi-joint movement. Heteronymous synergies
develop later, alongside faster limb movement, and their growth carries
kinematic information beyond the local afferent. This predicts that changing
the musculoskeletal embodiment, which changes how spontaneous activity moves
the limb, should reshape the synergist--antagonist balance of the learned
connectivity.



\begin{figure*}[t!]
  \centering
  \begin{overpic}[width=\linewidth]{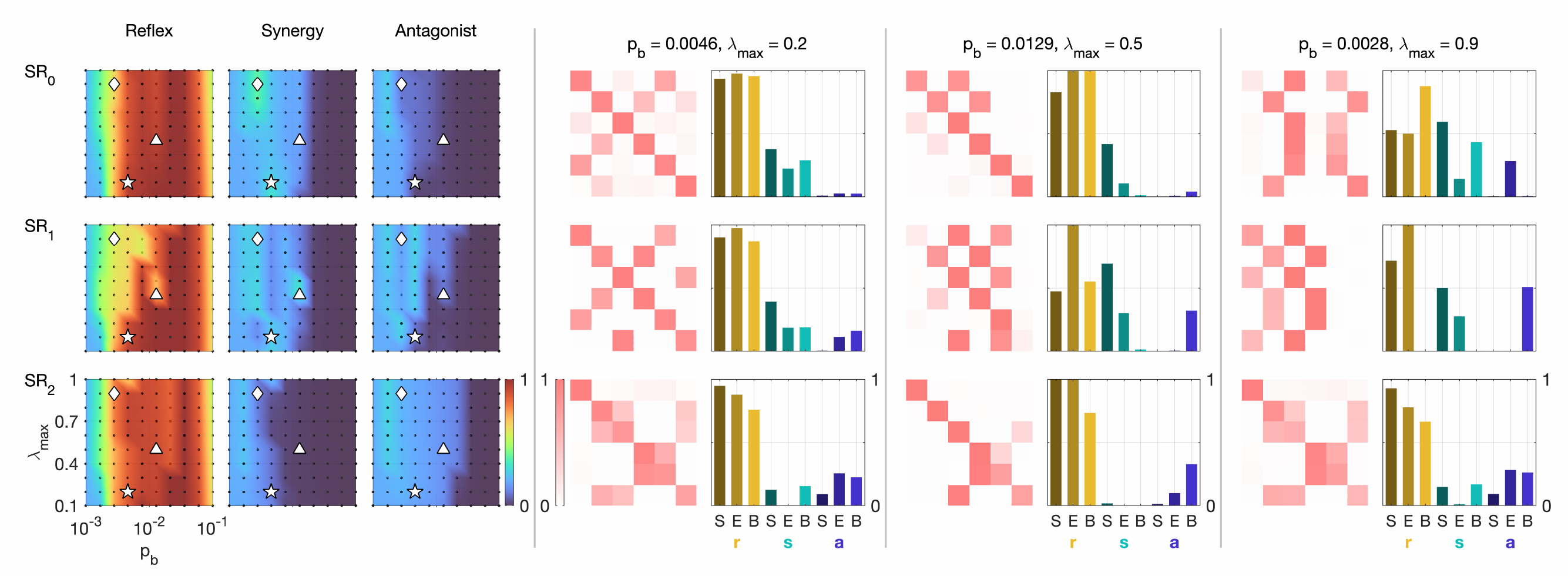}
    {\footnotesize
    \put(4,34){\sflabel{A}}
    \put(35,34){\normalsize\ding{73}}}
    \put(58,34){$\triangle$}
    \put(80,34){$\Diamond$}
  \end{overpic}
\caption{\textbf{Embodiment shapes the self-organized spinal circuit.}
  \sflabel{A} Mean reflex (homonymous), synergy (heteronymous synergist), and
  antagonist motif strength of the learned Ia$\to$MN connectivity across a
  sweep of the per-step babble probability $p_{\mathrm b}$ ($x$, log scale) and
  the developmental strength ceiling $\lambda_{\max}$ ($y$), from the motif
  regression ($\kappa=5$; one simulation per grid point, marked by dots). Rows
  are three muscle strength ratios (shoulder:elbow:biarticular), with total
  strength held fixed: a mammal-inspired gradient SR$_0$~=~1:0.5:0.25, a
  strong-biarticular variant SR$_1$~=~1:0.25:0.5, and a biarticular-dominant
  ``mutant'' SR$_2$~=~0.5:0.5:0.75, all with proximal:distal mass ratio 2:1. In
  every body, synergist and antagonist connections are absent at high babble
  probability. \emph{Right Columns}: final weight matrix $W$ (rows,
  motoneurons; columns, Ia afferents; muscles SF, SE, EF, EE, BF, BE) and motif
  spectrum---reflex ($r$), synergy ($s$), and antagonist ($a$) coefficients by
  muscle group (S, E, B)---at three operating points marked on every panel.
  (\ding{73}) At the canonical point, SR$_0$ and SR$_1$ self-organize an
  ordered spectrum, $R>S>A$, whereas in the mutant synergy is weak and
  antagonist coupling exceeds it. ($\triangle$) At higher babble probability
  and strength, biarticular synergy is lost in every body and synergy is
  concentrated in the shoulder pool; SR$_1$ also develops biarticular
  antagonist coupling at the expense of its shoulder and biarticular reflexes,
  and the mutant loses synergy entirely, retaining only reflexes and
  biarticular antagonist coupling. ($\Diamond$) Sparse but strong activity
  yields body-specific defects: SR$_0$ develops strong synergy but incomplete
  shoulder and elbow reflexes, with elbow antagonist coupling; SR$_1$ loses its
  biarticular reflex, replaced by biarticular antagonist coupling; the mutant's
  spectrum resembles its canonical one. Overall, babble probability sets
whether an ordered circuit forms, while embodiment and strength shape its fine
structure.} \label{fig:sweep}
\end{figure*}

\smallskip \noindent\textbf{Embodiment determines what connectivity
self-organizes.}\label{sec:embodiment} If synergies develop only after
sufficient limb movement, then both the statistics of spontaneous activity and
the mechanics of the body, which together set how that activity is converted
into movement, should shape the connectivity that self-organizes. We tested
this by sweeping the babble probability $p_{\mathrm b}$ against the
developmental strength ceiling $\lambda_{\max}$ for three strength ratios, with
total strength held fixed (Fig.~\ref{fig:sweep}): a mammal-inspired gradient
(SR$_0$~=~1:0.5:0.25; strongest shoulder, weakest biarticular), a
strong-biarticular variant (SR$_1$~=~1:0.25:0.5), and a biarticular-dominant
``mutant'' (SR$_2$~=~0.5:0.5:0.75). The limited range of mass distributions
that we examined had only small effects on these trends, which are shown in the
Supporting Information (Fig.~\ref{fig:si-sweep-mr1}, Table~\ref{tab:si-motif}).
To separate the operating point from embodiment, we compared all bodies at
three fixed $(p_{\mathrm b},\lambda_{\max})$ points: the canonical point of
Fig.~\ref{fig:motif-regression} (\ding{73}), a higher-babble, higher-strength
point ($\triangle$), and a sparse, high-strength point ($\Diamond$).

Babble probability governed whether an ordered circuit formed, and it did so
similarly in every body: the sweeps are organized mainly along $p_{\mathrm b}$,
and at high babble probability synergist and antagonist connections were absent
in all three bodies, leaving reflexes alone. At the canonical point
(\ding{73}), the mammal-inspired and strong-biarticular bodies self-organized
the ordered spectrum $R>S>A$, as in Fig.~\ref{fig:motif-regression}.

At the same operating point, embodiment shaped the learned connectivity.
Homonymous reflexes formed in every body, but the mutant, whose strongest
muscles are biarticular, did not reach the ordered spectrum: its synergy was
weak, and its antagonist coupling, concentrated on the elbow and biarticular
pairs, was stronger than its synergy (Fig.~\ref{fig:sweep}, \ding{73}).

Away from the canonical point, the biarticular muscles were the most sensitive
to embodiment. At higher babble probability and strength ($\triangle$),
biarticular synergy was lost in every body, and the two bodies with strong
biarticular muscles (SR$_1$, SR$_2$) developed biarticular antagonist coupling
instead, which the mammal-inspired body did not. Sparse but strong activity
($\Diamond$) produced body-specific defects: the mammal-inspired body developed
strong synergy but incomplete shoulder and elbow reflexes, whereas the
strong-biarticular body lost its biarticular reflex entirely, replaced by
biarticular antagonist coupling.

Together, these results separate two influences on self-organization. The
statistics of spontaneous activity set whether an ordered circuit forms at all,
a gate common to every body. Embodiment then shapes the connectivity realized:
under a proximal-to-distal strength gradient, biarticular connections remain
synergist or decay, whereas strong biarticular muscles convert them into
antagonist coupling away from the canonical point.
This refines the prediction from the mechanism analysis
(Fig.~\ref{fig:gc-partialR2}): embodiment reshapes the synergist--antagonist
balance of the learned spectrum, and it does so most strongly for the muscles
that span both joints.

\section*{Discussion} \label{sec:discussion}

\smallskip \noindent\textbf{Functional self-organization vs. genetic
predestination.} Experiments that have been interpreted in favor of genetic
predestination are not as conclusive as is often assumed. Loss of
muscle-specific Ia--MN connectivity following fetal paralysis was thought to
rule out functional self-organization~\cite{mendelson1991specific} but
overlooked limitations of the curare methodology, which probably does not block
fusimotor activity~\cite{proske1995sensitivity,enander2022model1}.

Similarly, the absence of preferential homonymous Ia--MN synapses in FoxP1
mutant mice with functional but randomly positioned motoneurons was explained
as the absence of spatial rather than functional cues~\cite{arber2012motor},
but this overlooks the likely concomitant loss of muscle-specific twitches that
arise from gap junctions between adjacent
motoneurons~\cite{personius2007reduced}. Furthermore, the spatial organization
of normal mammalian motor nuclei does not extend to the dendrites on which
Ia--MN synapses are located, which are highly overlapped among functionally
distinct motoneurons~\cite{keirstead1983dendritic,vanner1984dendritic}.

Greatly weakened but homonymously preferential Ia--MN synapses were demonstrated
in ErbB2$\Delta$ mutant mice that had mutant spindles with preserved Ia
afferents but only a single immature intrafusal
fiber~\cite{shneider2009functionally}. It is not known whether that intrafusal
fiber received fusimotor innervation or what Ia activity might have been
present during fetal development, either as a result of fusimotor activity or
mechanical stretch from spontaneous motor activity. If Ia activity was absent
and genetic markers in the spindle afferents were responsible for the
preferential connectivity, then the connectivity should have been preserved in
the FoxP1 mutants described above, but it was absent.

\smallskip \noindent \textbf{The body identifies its own controller.}
Spontaneous fetal movement is frequently described as undirected noise; our
results recast it as the exploratory probe of a self-identification process.
Sleep twitches and spontaneous movements have long been proposed to guide
sensorimotor development~\cite{blumberg2015developing,
petersson2003spontaneous}; our model specifies \emph{what} that guidance
teaches and shows the lesson to be body-dependent. In the language of system
identification, motor babble is the excitation signal, the musculoskeletal limb
is the plant, the Hebbian rule is the estimator, and the learned weight matrix
$W$ is the set of identified parameters. What can be identified is constrained
by what the body makes observable: only correlations that the limb actually
produces during movement are available to plasticity. This is morphological
computation in a concrete form~\cite{pfeifer2006body,
paul2006morphological}---the physical dynamics of the body supply the
regressors, and the spinal circuit reads them out.

This framing explains why the two motif classes are gated so differently.
Homonymous reflexes depend only on a muscle's own activation-coupled
(fusimotor) afferent and are identifiable while muscles are still weak;
heteronymous synergies depend on multi-joint kinematic coupling that develops
as spontaneous activity moves the limb faster
(Figs.~\ref{fig:gc-grid},~\ref{fig:gc-partialR2}). In identification terms,
estimating the proper weights, especially the cross-joint terms, requires
\emph{persistent excitation}: the input must drive multi-segmental motion
before those parameters can be estimated. A muscle that is too weak to move the
limb, or a babble that is too sparse to excite coordinated movement, fails this
condition. A babble that is too frequent fails differently, pruning the
cross-joint terms so that only the homonymous terms remain
(Figs.~\ref{fig:motif-regression},~\ref{fig:sweep}). The dependence of
synergies on induced musculoskeletal motion predicts that any manipulation
reducing multi-joint movement during development should selectively impair
heteronymous, but not homonymous, connectivity. The developmental experiment of
Mendelson and Frank should be repeated with an examination of heteronymous, not
just homonymous Ia--MN connectivity~\cite{mendelson1991specific}.

\smallskip \noindent \textbf{Stereotypy, individuality, and body-appropriate
circuits.} A long-standing puzzle is how spinal connectivity can be both
experience-dependent and highly stereotyped across
individuals~\cite{ladle2007assembly}. Our model offers a simple resolution: the
circuits are stereotyped because the bodies and their fetal behaviors are
stereotyped. Embodiment narrows the space of learnable correlations, canalizing
self-organization toward a narrow band of outcomes in response to the
randomness of the babble that drives it. A corollary is that residual
individual variation in connectivity should track individual variation in body
mechanics rather than being unstructured noise.

The same logic reframes what counts as an ``abnormal'' circuit. When we altered
the body---strengthening the biarticular muscles---self-organization produced a
different connectivity, in which biarticular connections became antagonist
rather than synergist and, in the mutant, antagonist coupling exceeded synergy
(Fig.~\ref{fig:sweep}). It is tempting to read this as maladaptive, but the
more intriguing interpretation is that the circuit remains matched to its body:
self-organization identifies the controller appropriate to the plant it is
given. Whether such body-specific circuits are in fact functionally appropriate
is a question our connectivity analysis cannot answer, as it requires testing
whether the learned circuit improves motor control of its own body. That is the
aim of ongoing work pairing these self-organized circuits with motor learning
across a range of tasks. The next step is to turn the present descriptive
result into a functional hypothesis: the reflex architecture that emerges from
a body is the one best suited to control it. The same principle suggests a
design strategy for embodied machines whose dynamics are likely to be different
from extant vertebrates. Rather than hand-specifying textbook connectivity
patterns, it may be possible to put such a machine through the equivalent of
fetal development and let the same learning rule discover the appropriate
connectivity~\cite{pfeifer2006body}.

\smallskip \noindent \textbf{Developmental order and critical periods.} In the
model, homonymous reflexes form first and heteronymous synergies later, as
strength and limb movement mature
(Figs.~\ref{fig:motif-regression},~\ref{fig:gc-grid}). This predicts the same
temporal order in development. This ordering is testable with postnatal
neurophysiology in animal models~\cite{mears1997formation,
shneider2009functionally}. Our model is silent, however, on whether the
plasticity that builds these circuits persists after fetal development.
Physiologically, both the parameters and perhaps the very existence of
plasticity likely change with age, raising the possibility of \emph{critical
periods} during which the body's correlational structure is read out and after
which the circuit is largely fixed. If so, the developmental window over which
strength and movement statistics mature would itself help set the final
connectivity. This bears on adult reorganization: should strength ratios change
later in life through injury, disease, or training, our framework predicts that
any reorganization is limited both by residual plasticity and by the
correlations the altered body can supply~\cite{wolpaw2007spinal}---with
implications for rehabilitation after stroke or spinal cord injury.

\smallskip \noindent \textbf{Variation across bodies, species, and disease.} If
connectivity is set by the body's strength layout, bodies that differ in that
layout should differ correspondingly in their heteronymous projections. Our
results point to the biarticular muscles as the most sensitive to this layout:
when they are strong, their connections shift from synergist to antagonist
(Fig.~\ref{fig:sweep}). Within humans, muscle-volume ratios differ between the
upper and lower limbs, predicting limb-specific synergy patterns. Across
species the contrasting forelimb and hindlimb mechanics of quadrupeds and
bipeds, and the unusual morphologies of animals such as bats and cetaceans,
should be reflected in correspondingly unusual spinal organization. The same
prediction extends to disease, with an important caveat. Disorders that
selectively alter proximal-versus-distal muscle strength, such as the muscular
dystrophies, should reshape the correlational structure available for synergy
formation and thus the learned circuit. Cerebral palsy is a tempting target,
given that movement quality in early infancy predicts later
outcome~\cite{einspieler2004prechtl}, but the link is not straightforward:
cerebral palsy typically reflects perinatal injury, whereas the circuits
studied here may mature largely before birth. More generally, some form of
persistent, activity-dependent synaptic plasticity may account for the
well-documented childhood progression of musculoskeletal and reflex pathology:
as an atypical body develops, self-organization should track it. Spastic
reflexes involving antagonist cocontraction, such as those that emerged in some
of our unphysiological models, are a common and progressive feature in children
with cerebral palsy~\cite{myklebust1982reciprocal}, who tend also to develop
progressive musculoskeletal pathology~\cite{graham2003musculoskeletal}.
Following the argument above, the resulting circuit may be appropriate to that
body rather than simply deficient. Distinguishing genuinely maladaptive
connectivity from body-appropriate adaptation may itself be a useful reframing
for developmental motor disorders.

\smallskip \noindent \textbf{Limitations and outlook.} Our model is
deliberately reduced from an anatomically complete limb to facilitate
demonstration of organizing principles. The connectivity comparison is
qualitative, based on a conserved rank ordering of motifs, rather than a
quantitative fit to any species. Most importantly, we have characterized the
\emph{structure} of the self-organized circuits but not their \emph{function}.
Establishing that body-matched connectivity confers a control advantage, and
mapping how that advantage depends on task and body, is an important next step.
Self-organization offers an explanation for how spinal circuits arise and the
specific motifs that have been identified experimentally in adult animals.
Hebbian self-organization also provides a plausible starting point to solve
Baldwin's conundrum of how an animal with a useful musculoskeletal mutation can
learn to survive and propagate that
mutation~\cite{baldwin1896new,baldwin1897organic,simpson1953baldwin}.

\subsubsection*{Methods} \label{sec:methods}
\small

\noindent The model is a closed sensorimotor loop (Fig.~\ref{fig:model}):
spinal motoneurons drive a musculoskeletal limb, proprioceptive afferents sense
the resulting movement, a plastic Ia--motoneuron projection $W$ couples them, and
a spontaneous drive $\mathbf{d}$ perturbs the loop. The embodiment
parameters (strength ratio SR, mass ratio MR) and the Hebbian plasticity rule are
defined in the Introduction; here we specify the remaining components and give
all parameter values (Table~\ref{tab:params}). Signals update at the simulation
rate $f_{\mathrm s} = 1/\Delta t = 200$~Hz; the spinal network and descending drive are
evaluated at every simulation step.

\medskip \noindent\textbf{Musculoskeletal plant.} The limb is a planar
two-joint arm (shoulder $q_1$, elbow $q_2$) actuated by six
muscles---monoarticular shoulder and elbow flexors and extensors (SF, SE, EF,
EE) and two biarticular muscles (BF, BE)---simulated as a rigid-body system,
$M(\mathbf{q})\ddot{\mathbf{q}} + C(\mathbf{q},\dot{\mathbf{q}})
\dot{\mathbf{q}} = J_t^\top\mathbf{F}$, and integrated with
MuJoCo~\cite{todorov2012mujoco}. Each muscle is a MuJoCo muscle actuator with
force--length--velocity properties (Fig.~\ref{fig:model}B; geometry and
curves in Fig.~\ref{fig:si-plant} and Table~\ref{tab:si-geometry}); given a neural
command (activation) $a\in[0,1]$, the $i^\text{th}$ muscle produces
\[
  F_i = \lambda(t)\,\rho_i\,F_{\mathrm{ref}}\,f_{\mathrm m}(a_i,L_i,\dot{L}_i),
\]
where $F_{\mathrm{ref}}=240$~N is a fixed reference force, $\rho_i$ the muscle's
relative strength (set by SR), $\lambda(t)$ the developmental gain, $L_i$ the
normalized muscle length, and $f_{\mathrm m}$ MuJoCo's force--length--velocity function
with operating range $[0.6,1.4]$ optimal lengths~\cite{todorov2012mujoco}.
Embodiment is set by the relative strengths $\rho_i$ (SR) and the
proximal:distal segment masses (MR; 2:1 or 1.5:1.5~kg), with total strength and
total mass held fixed. Gravity is off, representing buoyancy in the amniotic
fluid; joint damping is zero, and the limb moves through a viscous medium
($0.01$~Pa$\cdot$s) that supplies mass-proportional damping.
Each simulation comprises 2000 episodes of 90~s
(36M steps, 50~h); limb state is continuous across episodes. Muscle force
capacity matures through $\lambda(t)$, which is held constant within each
episode and increased in equal steps between episodes from
$\lambda_{\min}=0.01$ to $\lambda_{\max}$ over the first 1000 episodes (25~h),
then held at $\lambda_{\max}$.

\smallskip \noindent\textbf{Proprioceptive afferents.} Each muscle carries a
spindle model that maps its normalized length $L$, velocity $\dot{L}$, and own
activation $a$ to secondary (II) and primary (Ia) afferent signals,
\[
  x_{\mathrm{II}} =
  \Big[\tfrac12\big(g_{\mathrm L}(L+b_{\mathrm{II}}) + a\big)\Big]_0^1 ,
\]
\[
  x_{\mathrm{Ia}} =
  \Big[\tfrac12\big((c_{\mathrm{sat}} + \log_{10}(a+0.1))\,g_{\mathrm v}\,\dot{L}
        + g_{\mathrm{II}}\,x_{\mathrm{II}} + a\big)\Big]_0^1 
\]
where $[ \cdot ]_0^1$ denotes the clip operator. The dependence on the muscle's
own activation $a$ realizes the beta fusimotor coupling that defines
$x_{\mathrm{Ia}}=f_{\mathrm{Ia}}(a,L,\dot{L})$: when a
muscle's motoneuron is active its spindle is excited rather than silenced, and
the velocity sensitivity is itself activation-gated through the $\log_{10}$
term.

\smallskip \noindent\textbf{Spinal motoneurons.}
Expanding the compact form given in~\eqref{eq:mn}, the motoneuron pool combines
the reflex current $W\mathbf{x}_{\mathrm{Ia}}$ with the descending drive
$\mathbf{d}$, applies a divisive (shunting) normalization by the total synaptic
conductance, low-passes the result with the membrane filter
$\Psi_{\tau_{\mathrm s}}$, and saturates:
\[
  \mathbf{u} = W\mathbf{x}_{\mathrm{Ia}} + \mathbf{d},\qquad
  \mathbf{v} = \frac{\mathbf{u}}{\,s_0 + s_1\,|\mathbf{u}|\,},\qquad
  \mathbf{x}_{\mathrm{MN}} = \big[\,\Psi_{\tau_{\mathrm s}}(\mathbf{v})\,\big]_0^1 .
\]
Motoneuron activity is converted to the muscle command by a clipped affine map,
\[
  \mathbf{a} = \big[\,g_{\mathrm m}(\mathbf{x}_{\mathrm{MN}} + b_{\mathrm m})\,\big]_0^1 ,
\]
which is sent to the actuators; MuJoCo then applies its own first-order activation
dynamics ($\Psi_{\mathrm a}$) before generating force, closing the loop.

\smallskip\noindent\textbf{Spontaneous motor babble.}
Each motoneuron $i$ receives an independent descending drive $d_i(t)$ generated
as a renewal process. While no event is active, an onset occurs at each step
with probability $p_{\mathrm b}$ (a Bernoulli trial), so the waiting time from
the end of one event to the next onset is geometric (exponential in continuous
time). On onset, an amplitude $A\sim\mathcal U(0,1)$, a duration $T\sim\mathcal
U[T_{\min},T_{\max}]$, and an onset-smoothing constant $K_{\mathrm
r}\sim\mathcal U[K_{\mathrm r,\min},K_{\mathrm r,\max}]$ are drawn; $d_i$ is
then a unit pulse over $[t_0,t_0+T]$, scaled by $A$ and smoothed by a
first-order filter with constant $K_{\mathrm r}$. In the Hebbian-learning runs
studied here, babble is the sole external drive, and $p_{\mathrm b}$ is a
primary swept parameter.

\smallskip\noindent\textbf{Hebbian plasticity and temporal filters.}
The plastic matrix $W$ ($6\times6$,
$\mathbf{x}_{\mathrm{Ia}}\!\to\!\mathbf{x}_{\mathrm{MN}}$) is initialized with
small random weights ($w_{ij}\sim\mathcal N(0.2,0.01)$) and updated by the
eligibility-filtered, soft-bounded rule in \eqref{eq:hebbian}, with presynaptic
signal $\mathbf{x}_{\mathrm{Ia}}$ and postsynaptic signal
$\mathbf{x}_{\mathrm{MN}}$: the eligibility $\ell_{ij}$ integrates the product
of the presynaptic afferent, the current weight, and the postsynaptic deviation
$\kappa\,\Psi_{\tau_{\mathrm H}}(x_{\mathrm{MN},i})-\bar{x}_{\mathrm{MN},i}$;
the soft bound $c_{ij}$ keeps $w_{ij}\in[0,1]$; and the rate
$\eta_i=\bar{x}_{\mathrm{MN},i}^4\,\eta$ gates plasticity by tonic motoneuron
activity. All temporal filters are first-order discrete IIR (exponential moving
averages),
\[
  y[n] = K\,x[n-1] + (1-K)\,y[n-1],\qquad \tau \approx -\Delta t/\ln(1-K),
\]
realizing the membrane filter $\Psi_{\tau_{\mathrm s}}$ (constant $K_{\mathrm
s}$), the eligibility/learning filter $\Psi_{\tau_{\mathrm L}}$ ($K_{\mathrm
L}$), and the slow tonic mean $\bar{x}_{\mathrm{MN}}$ ($K_{\bar x}$); the
babble onset uses the same form with constant $K_{\mathrm r}$. The phasic
filter $\Psi_{\tau_{\mathrm H}}$ is a first-order Butterworth high-pass with
normalized cutoff $f_{\mathrm c}$.

\smallskip\noindent\textbf{Motif decomposition.}
To characterize the learned connectivity we approximate $W$ as a weighted sum of
fixed connectivity motifs. We define nine binary motif templates spanning
homonymous reflex, heteronymous synergist, and antagonist connections, each
resolved by muscle class (shoulder S, elbow E, biarticular B). Stacking the
vectorized templates as columns of a matrix $B$, the motif coefficients are the
least-squares projection
\[
  \mathbf{c} = B^{+}\,\mathrm{vec}(W)
   = (r_{\mathrm S},r_{\mathrm E},r_{\mathrm B},\,
      s_{\mathrm S},s_{\mathrm E},s_{\mathrm B},\,
      a_{\mathrm S},a_{\mathrm E},a_{\mathrm B})^{\!\top},
\]
where $B^{+}$ is the Moore--Penrose pseudoinverse. Averaging within type gives the
global reflex $R$, synergy $S$, and antagonist $A$ strengths.
Applying the projection to $W(t)$ across learning
yields the motif time courses and final spectra
(Figs.~\ref{fig:motif-regression},~\ref{fig:sweep}).

\smallskip\noindent\textbf{Granger causality analysis.} To test whether limb
kinematics predict weight growth beyond local neural activity, we ran an
event-triggered Granger causality analysis for each homonymous and synergist
connection $(i,j)$. Antagonist connections, which all decay, were not tested.

\emph{Events.} From the weight trace $w_{ij}(t)=[W(t)]_{ij}$ we form the rate
$\dot w_{ij}$ and take local peaks of its positive part,
$\dot w^{+}=\tfrac12(\dot w_{ij}+|\dot w_{ij}|)$. Peaks are retained while the
weight lies below 90\% of its maximum, $w_{ij}<0.9\max_t w_{ij}$, which
excludes the plateau of saturated weights; the same rule is applied to every
connection. The rate $\dot w_{ij}$ was computed from the learning rule at each
step rather than by differencing the stored weights, whose per-step increments
approach single-precision resolution. Let $\{t_k\}$ be the retained event times.

\emph{Features.} For each event we take a strictly pre-event window of
$L=\tau_{\mathrm{lag}}/\Delta t=80$ samples ($\tau_{\mathrm{lag}}=0.4$~s),
$\mathcal H_k=\{t_k-L,\dots,t_k-1\}$, chosen to contain the longest peak lag
from afferent to weight change (0.38~s; Fig.~\ref{fig:si-ia-lag}), and use the
lagged samples of the weight rate $\dot w_{ij}$ (history), the presynaptic
afferent $x_{\mathrm{Ia},j}$, and the joint angles and velocities
$(\mathbf{q},\dot{\mathbf{q}})$ as regressors. The response (regression target)
is the peak rate $y_k=\dot w_{ij}(t_k)$.

\emph{Restricted and full models.} We compare, via ordinary least squares for
each $(i, j)$, a restricted model $m_0:\{\mathrm{Ia},h\}$ against a full
(unrestricted) model $m_1:\{\mathrm{Ia},\mathbf{k},h\}$ that adds the
kinematics. Because $x_{\mathrm{Ia},j}$ already encodes muscle $j$'s own length
and velocity, kinematic information beyond it reflects movement not captured
locally, such as coupled motion of the other joint. The history term $h$ is
essential because $\dot w_{ij}$ is strongly autocorrelated; without it, slow
drift in the weight change would be attributed to the kinematics.

\emph{Statistics.} Following standard practice in Granger causality, for each
$(i, j)$ we report the corresponding $F$-statistic,
$F=\big[(\mathrm{SSE}_0-\mathrm{SSE}_1)/(\mathrm{df}_0-\mathrm{df}_1)\big]/(\mathrm{SSE}_1/\mathrm{df}_1)$,
and its associated $p$-value, Bonferroni-corrected across the 18 tested
connections. Because events are autocorrelated and peak-selected, $p$-values
are anti-conservative. Synergist connections also contribute more events than
homonymous ones (5.3--9.1$\times$10$^4$ vs.\ $\sim$1.9$\times$10$^4$ per
connection), which increases their power; we therefore interpret the separation
between the two classes qualitatively.

\begin{table}[!t]
  \centering
  \footnotesize
  \caption{\textbf{Model and analysis parameters.} Swept parameters list the
  range used; the values used in each figure are given in the figure.}
  \label{tab:params}
  \begin{tabular}{l l l}
    \toprule
    Symbol & Description & Value \\
    \midrule
    \multicolumn{3}{l}{\emph{Plant and simulation}}\\
    $\Delta t$ & simulation timestep & 5~ms (200~Hz)\\
    --- & episode length; episodes per simulation & 90~s; 2000 (50~h)\\
    --- & gravity; joint damping & 0; 0\\
    --- & medium viscosity & 0.01~Pa$\cdot$s\\
    $F_{\mathrm{ref}}$ & reference peak isometric force & 240~N\\
    --- & muscle operating range & $[0.6,1.4]\,\ell_0$\\
    $v_{\max}$ & maximum shortening velocity & $1.5\,\ell_0\,\mathrm{s}^{-1}$\\
    $f_{\mathrm p}^{\max}$ & passive-force scale ($1.5f_{\mathrm p}^{\max}$ at $1.4\,\ell_0$) & 1.3\\
    $f_{\mathrm v}^{\max}$ & max lengthening force gain & 1.2\\
    \midrule
    \multicolumn{3}{l}{\emph{Embodiment}}\\
    SR & strength ratio S:E:B & $1{:}0.5{:}0.25,\ 1{:}0.25{:}0.5,\ 0.5{:}0.5{:}0.75$\\
    MR & mass ratio proximal:distal & $2{:}1,\ 1.5{:}1.5$~kg\\
    \midrule
    \multicolumn{3}{l}{\emph{Developmental scaling}}\\
    $\lambda_{\min}$ & initial strength gain & 0.01\\
    $\lambda_{\max}$ & strength ceiling (swept) & $[0.1,\ 1]$\\
    --- & strength ramp & linear, stepped per episode over episodes 1--1000\\
    \midrule
    \multicolumn{3}{l}{\emph{Spindle (afferents)}}\\
    $g_{\mathrm L}$ & II length gain & 1.25\\
    $b_{\mathrm{II}}$ & II length bias & $-0.2$\\
    $g_{\mathrm{II}}$ & Ia$\leftarrow$II gain & 0.2\\
    $c_{\mathrm{sat}}$ & Ia velocity-saturation constant & 1.5\\
    $g_{\mathrm v}$ & Ia velocity gain & 1\\
    \midrule
    \multicolumn{3}{l}{\emph{Motoneuron and muscle command}}\\
    $s_0$ & static leak (shunt) & 0.6\\
    $s_1$ & dynamic leak (shunt) & 1\\
    $K_{\mathrm s}$ & membrane filter constant ($\Psi_{\tau_{\mathrm s}}$) & 0.3\\
    $g_{\mathrm m}$ & muscle-command gain & 1.11\\
    $b_{\mathrm m}$ & muscle-command bias & $-0.1$\\
    $W_0$ & initial weights & $\mathcal N(0.2,0.01)$\\
    \midrule
    \multicolumn{3}{l}{\emph{Motor babble}}\\
    $p_{\mathrm b}$ & onset probability per step (swept) & $[10^{-3},\ 10^{-1}]$\\
    $[T_{\min},T_{\max}]$ & pulse duration & $[0.05,\ 0.1]$~s\\
    $[K_{\mathrm r,\min},K_{\mathrm r,\max}]$ & onset-smoothing constant & $[0.5,\ 0.8]$\\
    \midrule
    \multicolumn{3}{l}{\emph{Hebbian plasticity}}\\
    $\kappa$ & Hebbian gain & 5 (also 1--4, 10)\\
    $\eta$ & learning rate & 0.1\\
    $K_{\bar x}$ & tonic-mean filter constant ($\bar{x}_{\mathrm{MN}}$) & $4\times10^{-5}$\\
    $K_{\mathrm L}$ & learning filter constant ($\Psi_{\tau_{\mathrm L}}$) & $10^{-3}$\\
    $f_{\mathrm c}$ & phasic high-pass cutoff ($\Psi_{\tau_{\mathrm H}}$, normalized) & $5\times10^{-4}$\\
    $[w_{\min},w_{\max}]$ & weight bounds (soft) & $[0,\ 1]$\\
    \midrule
    \multicolumn{3}{l}{\emph{Granger causality analysis}}\\
    $\tau_{\mathrm{lag}}$ & pre-event history window & 0.4~s (80 samples)\\
    --- & event gate & $w_{ij}<0.9\max_t w_{ij}$; positive $\dot w_{ij}$ peaks\\
    \bottomrule
  \end{tabular}
\end{table}

\smallskip\noindent\textbf{Simulation and implementation.} The neuromechanical
model is implemented in Python, using MuJoCo~\cite{todorov2012mujoco} for the
physics and PyTorch for the spinal network and plasticity; the motif
decomposition and Granger causality analysis are implemented in MATLAB. Parameter
sweeps cross $\lambda_{\max}\in\{0.1,0.2,\dots,1\}$ with ten log-spaced values
of $p_{\mathrm b}\in[10^{-3},10^{-1}]$ at $\kappa=5$, for six bodies (three SR
$\times$ two MR), with one simulation (fixed random seed) per grid point; the
Hebbian gain was additionally swept ($\kappa\in\{1,2,3,4,10\}$) for SR$_0$,
MR$_0$. Figs.~\ref{fig:gc-grid} and~\ref{fig:gc-partialR2} use a separate
simulation at the canonical point ($\lambda_{\max}=0.2$, $p_{\mathrm
b}=4.6\times10^{-3}$) logged at the full 200~Hz rate.

\subsection*{Data, Materials, and Software Availability} All simulation code,
configuration files, launch scripts, and analysis scripts are available at
\url{https://github.com/jonreal/deep-spine/tree/v0.1.0}. No data were
deposited: the simulations are deterministic, and every result can be
regenerated with the provided launch scripts, following the steps in the
repository README.

\subsection*{Author Contributions} J.R., E.N., and G.E.L. designed research;
J.R. performed research; J.R., E.N., and G.E.L. analyzed data; J.R., E.N., and
G.E.L. wrote the paper.

\subsection*{Competing Interests} The authors declare no competing interest.

\subsection*{Acknowledgments} This work was supported by start-up funds from
the University of California, Riverside, to J.R. and NSF Award 2239654 and NIH
Award R01-NS-148491, to E.N.

\subsection*{Use of Generative AI} Claude Code (Anthropic; Claude 3.7 Sonnet
through Claude Opus 5.5) was the primary AI tool, used to help write, run, and
debug simulation and analysis code, search the literature, and edit text.
ChatGPT (OpenAI; GPT-4o through GPT-6) was used occasionally to discuss
mathematical and conceptual background. The authors reviewed all outputs,
checked all citations, and take full responsibility for the content.

\small
\bibliographystyle{unsrtnat}
\bibliography{citations-pnas}

\clearpage
\input{si.tex}

\end{document}

%% file: si.tex
\beginsupplement
\section*{Supporting Information}

\subsection*{Extended model details}

\emph{Spindle normalization and calibration.} Muscle length $L$ and velocity
$\dot L$ enter the spindle model in MuJoCo's own muscle normalization: length in
units of the optimal fiber length $\ell_0$, and velocity in units of the
maximal shortening velocity, $\dot L = v/(\ell_0 v_{\max})$ with
$v_{\max}=1.5\,\ell_0\,\mathrm{s}^{-1}$, so that $\dot L\in[-1,1]$ for every
muscle regardless of its geometry. At the rest posture ($q=0$) every muscle
sits near its optimal length. With the velocity gain $g_{\mathrm v}=1$ and saturation
constant $c_{\mathrm{sat}}=1.5$ (Table~\ref{tab:params}), and neglecting the
length (II) contribution, a passive spindle ($a=0$) stretched at maximal
velocity reaches $x_{\mathrm{Ia}}=0.25$, and an active spindle ($a=0.22$)
shortening at $\dot L=-0.22$ is silent; the
velocity sensitivity thus rises with activation through the $\log_{10}$ term.

\emph{Muscle force.} Peak isometric force is specified directly,
$F_i=\lambda(t)\,\rho_i\,F_{\mathrm{ref}}$, rather than derived from limb
inertia, so changing segment masses leaves muscle strength unchanged; we
verified that peak forces are identical across all mass ratios used.

\begin{figure}[!ht]
  \centering
  \includegraphics[width=\linewidth]{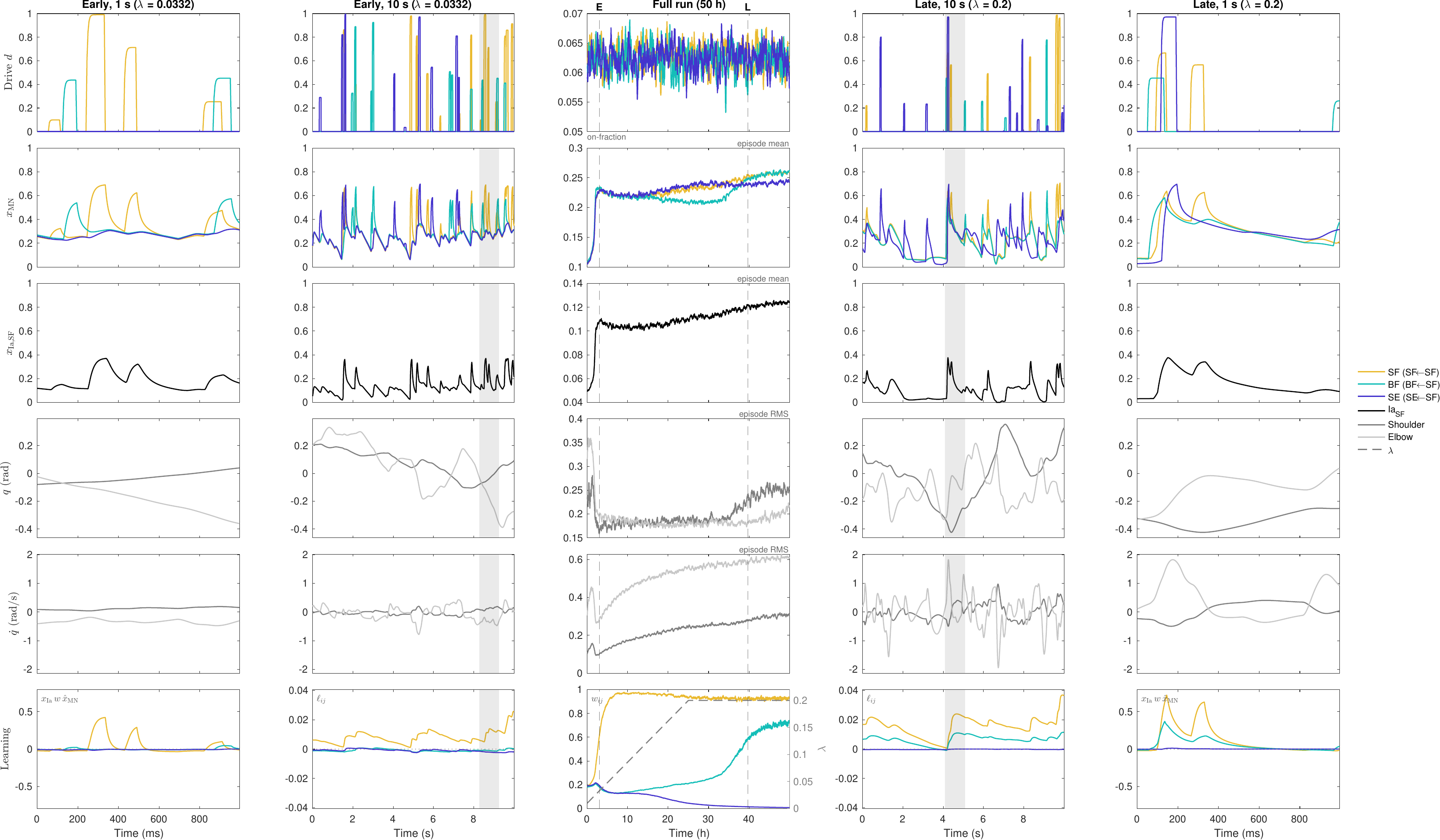}
  \caption{\textbf{Multi-scale time history of the canonical simulation}
  ($\kappa=5$, $\lambda_{\max}=0.2$, $p_{\mathrm b}=4.6\times10^{-3}$, SR$_0$, MR$_0$).
  Center: the full 50~h simulation; left and right: zooms to 10~s and 1~s
  windows early (E, 3.06~h, during reflex formation) and late (L, 39.7~h,
  during synergy growth) in development. Signals are shown for three example
  connections sharing the presynaptic afferent Ia$_{\mathrm{SF}}$: SF$\leftarrow$SF
  (reflex), BF$\leftarrow$SF (synergy), and SE$\leftarrow$SF (antagonist); drive
  and motoneuron traces are colored by the connection they target. On the full
  run, drive is the fraction of time on, neural signals are episode means,
  and kinematics are episode RMS values, all smoothed over 11 episodes; the
  learning row shows the weights with the strength ramp $\lambda(t)$ (dashed,
  right axis). On the 10~s panels the learning row shows the eligibility
  $\ell_{ij}$; on the 1~s panels, the instantaneous Hebbian term that
  $\ell_{ij}$ integrates. Windows were chosen by rule: the 10~s windows are
  centered on the steepest growth of the reflex (early) and of the synergy
  (late, searched between 30 and 40~h), and each 1~s window on the largest
  weight increase of that connection within its 10~s window (shaded).}
  \label{fig:si-timehistory}
\end{figure}

\emph{Spontaneous activity.} Each motoneuron receives babble pulses of
50--100~ms whose onset occurs with probability $p_{\mathrm b}$ per 5~ms step while no
pulse is active. At the canonical $p_{\mathrm b}=4.6\times10^{-3}$, each motoneuron is
driven about 6\% of the time (Fig.~\ref{fig:si-timehistory}), corresponding to
approximately 0.8 pulse onsets per second.

\emph{Motif decomposition.} The nine motif templates are disjoint binary masks
over the $6\times6$ connectivity matrix
(Fig.~\ref{fig:motif-regression}\sflabel{A}), so the least-squares projection
reduces to averaging: each coefficient ($r$, $s$, or $a$ for pool S, E, or B)
is the mean weight of the connections in its template. The global strengths
$R$, $S$, and $A$ are the means of the three pool coefficients of each type.

\emph{Learning signals.} Only the weights $W$ were logged during simulation.
The eligibility $\ell_{ij}$ and the instantaneous Hebbian term
$x_{\mathrm{Ia},j}\,w_{ij}\,(\kappa\,\Psi_{\tau_{\mathrm H}}(x_{\mathrm{MN},i})-\bar{x}_{\mathrm{MN},i})$
shown in Fig.~\ref{fig:si-timehistory} were recomputed offline from the logged
motoneuron and afferent activity and $W$, using the learner's filters and
parameters.

\subsection*{Modeling pitfalls}

Several modeling choices strongly affect the learned connectivity but are
easy to get wrong. We note those we encountered.

\emph{Muscle force scaling.} With MuJoCo's automatic force scaling
(\texttt{force="-1"}), peak isometric force is set from the effective inertia
each muscle drives. Specified relative strengths are then distorted (an
intended 1:0.5:0.25 shoulder:elbow:biarticular ratio became about
1:0.14:0.10 in our arm), and changing segment masses silently changes muscle force, coupling
strength and inertia. We therefore set peak force explicitly
($F_i=\lambda\rho_iF_{\mathrm{ref}}$) and verified that it is independent of
the mass ratio.

\emph{Damping and viscosity.} A single damping coefficient applied to both
joints leaves the shoulder underdamped and the elbow strongly overdamped,
because the two joints differ about ninefold in the inertia distal to each joint. Exploration is then
dominated by the shoulder, and elbow and biarticular correlations are
under-sampled. Mass-proportional damping avoids this, so we set joint damping
to zero and move the limb through a viscous medium instead. This medium is a
source of damping that arises from the whole uterine structure, not a model of
amniotic fluid, which is close to water in viscosity (about ten times less
viscous than our medium). The fluid's main mechanical effect, buoyancy, is
represented by omitting gravity.

\emph{Compensating one error with another.} An afferent velocity gain ten
times the physiological value at first appeared necessary to obtain usable Ia
signals from the elbow. The real cause was the weak, overdamped elbow above.
Because muscle velocity is normalized by optimal fiber length and maximum
shortening velocity, the Ia gains are independent of the plant, and the
physiological values ($g_{\mathrm v}=1$, $c_{\mathrm{sat}}=1.5$) work once the plant is
correct.

\begin{figure}[!ht]
  \centering
  \includegraphics[width=\linewidth]{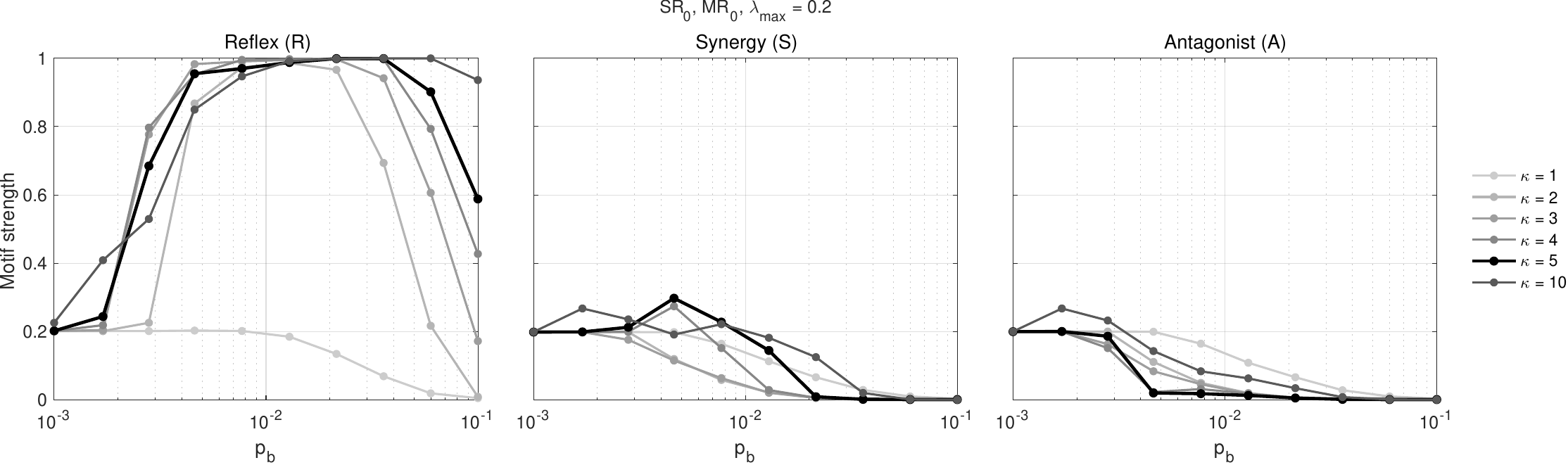}
  \caption{\textbf{Dependence of self-organization on the Hebbian gain.} Mean
  reflex, synergy, and antagonist strength versus babble probability for
  SR$_0$, MR$_0$ at $\lambda_{\max}=0.2$, for $\kappa=1$--5 and 10 (black,
  $\kappa=5$, used throughout). At $\kappa=1$, weights remain near their
  initial value (0.2) and decay at high babble probability; at $\kappa=2$--3,
  reflexes form but synergist and antagonist connections are not
  distinguished; synergy exceeding antagonist appears only for $\kappa\geq4$,
  within a window of babble probability.}
  \label{fig:si-kappa-lines}
\end{figure}

\emph{Hebbian gain.} With $\kappa=1$, weights change little over the
developmental period (Fig.~\ref{fig:si-kappa-lines}).

\emph{Numerical precision in analysis.} Weights stored in single precision
change by only a few units of the last decimal place per step. Rates obtained
by differencing stored weights are therefore quantized, which distorted our
Granger causality analysis until we computed rates from the learning rule.

\subsection*{Supporting movies}

\noindent\textbf{Movie S1. Full simulation.} The entire 50~h canonical
simulation at 2000$\times$ real time (90~s). Top left, the Ia--MN weight matrix
(white to red, 0 to 1); bottom, the current strength $\lambda(t)$; top, episode
number. Side traces are not interpretable at this speed; see Movies S2--S5.

\smallskip\noindent\textbf{Movie S2. Onset of development.} The first 20~s of the
canonical simulation (SR$_0$, MR$_0$, $\kappa=5$, $\lambda_{\max}=0.2$,
$p_{\mathrm b}=4.6\times10^{-3}$), in real time. Spontaneous babble drives the arm.
Top left, the Ia--MN weight matrix (white to red, 0 to 1); left and right,
the last 1~s of learning, neural, and kinematic signals; bottom, the current
strength $\lambda(t)$; top, episode and time within the episode.

\smallskip\noindent\textbf{Movie S3. Reflex formation.} 20~s centered on the
early window of Fig.~\ref{fig:si-timehistory} (3.06~h).

\smallskip\noindent\textbf{Movie S4. Synergy growth.} 20~s centered on the late
window of Fig.~\ref{fig:si-timehistory} (39.7~h).

\smallskip\noindent\textbf{Movie S5. Mature circuit.} The last 20~s of the
canonical simulation (50~h).

\subsection*{Supporting figures and tables}

\begin{figure}[!ht]
  \centering
  \includegraphics[width=\linewidth]{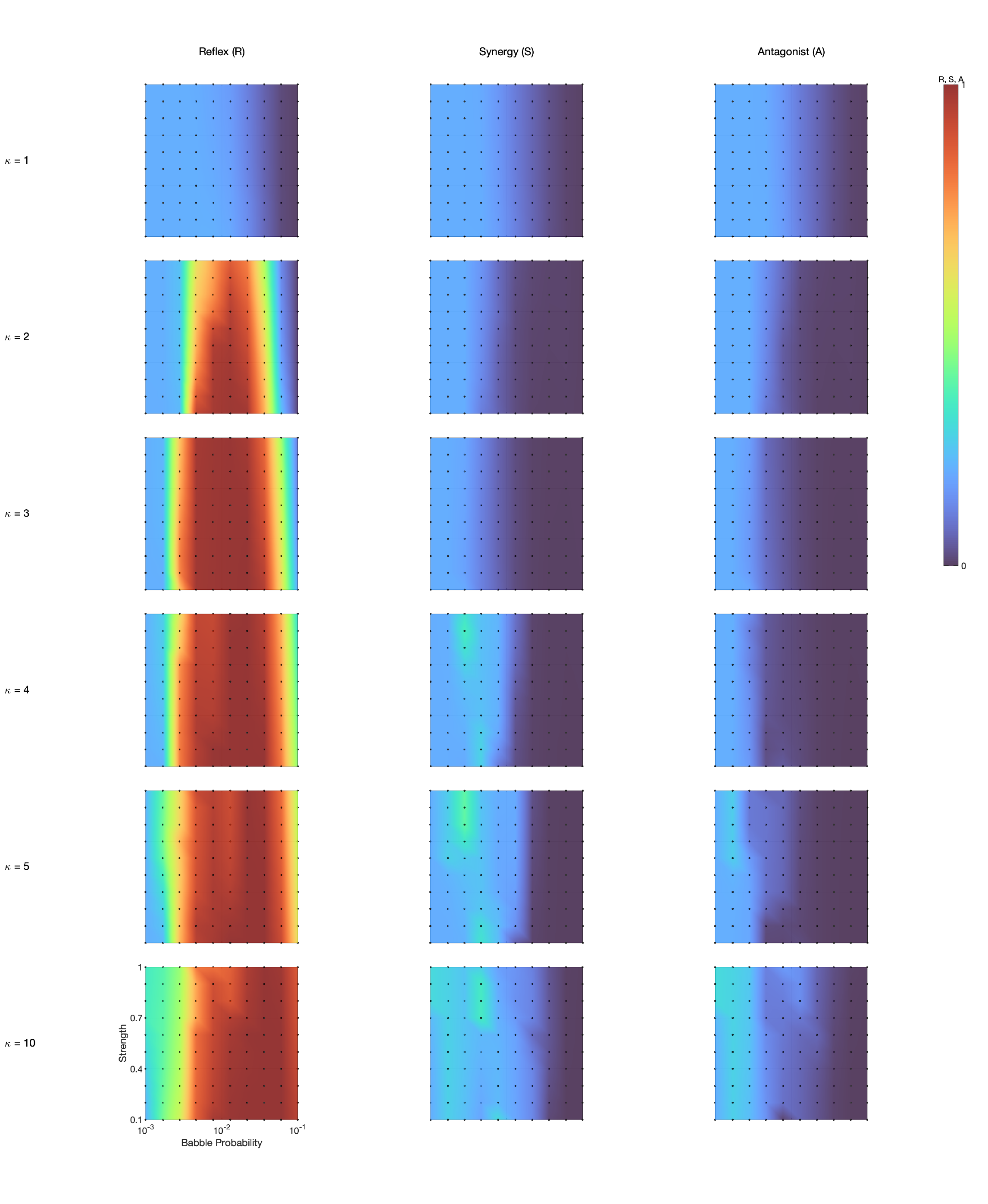}
  \caption{\textbf{Hebbian-gain progression across the full sweep}
  (SR$_0$, MR$_0$). Rows: $\kappa=1$--5 and 10. Columns: mean reflex, synergy,
  and antagonist strength over babble probability
  ($x$, log scale) and strength ceiling $\lambda_{\max}$ ($y$); dots mark
  simulated grid points.}
  \label{fig:si-kappa-heatmaps}
\end{figure}

\begin{figure}[!ht]
  \centering
  \begin{overpic}[width=\linewidth]{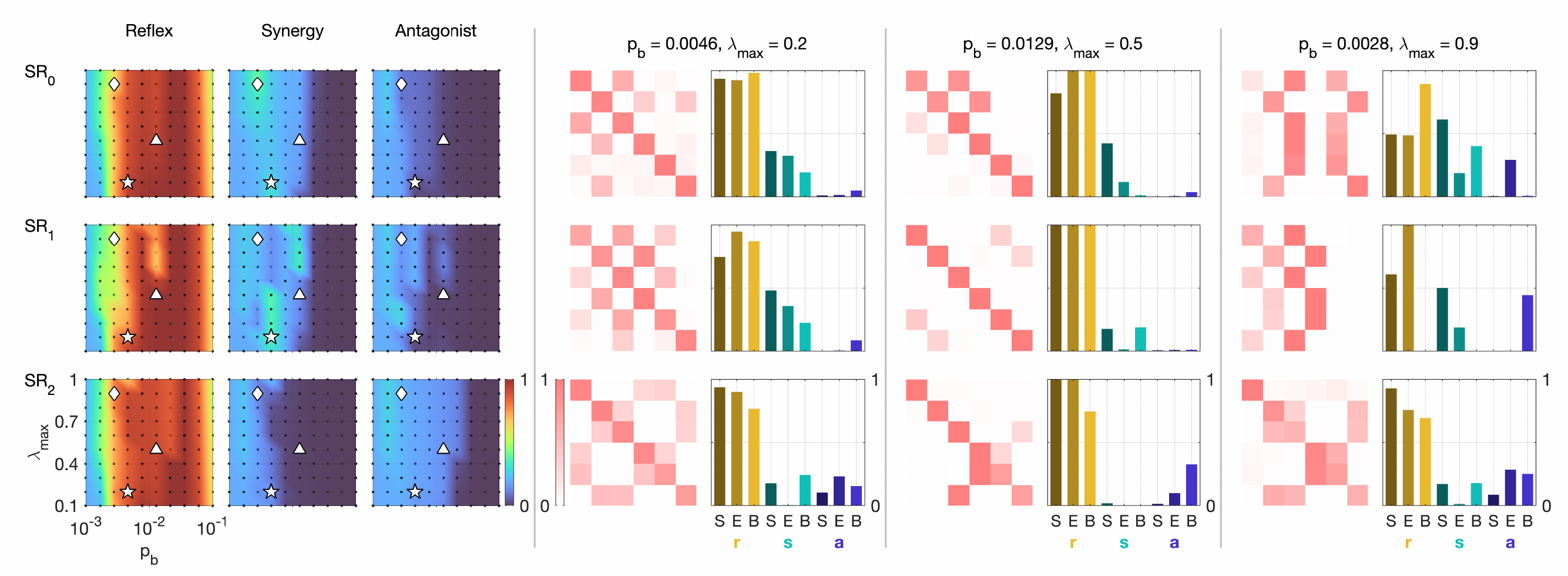}
    \put(35,34){\normalsize\ding{73}}
    \put(58,34){$\triangle$}
    \put(80,34){$\Diamond$}
  \end{overpic}
  \caption{\textbf{Embodiment sweep for a uniform mass distribution.} As
  Fig.~\ref{fig:sweep}, for the same three strength ratios with
  proximal:distal mass ratio MR$_1$~=~1.5:1.5. At the canonical point
  (\ding{73}) and the sparse, strong point ($\Diamond$), the learned spectra
  closely match those of MR$_0$ (Table~\ref{tab:si-motif}). At the
  higher-babble point ($\triangle$), SR$_1$ differs: it retains all
  homonymous reflexes and some biarticular synergy without developing
  biarticular antagonist coupling. This point lies on a sharp transition in
  the SR$_1$ sweep (Reflex column), where single simulations are sensitive to
  small changes in the body.}
  \label{fig:si-sweep-mr1}
\end{figure}

\begin{table}[!ht]
  \centering
  \footnotesize
  \caption{\textbf{Motif coefficients at the operating points of
  Fig.~\ref{fig:sweep}} for all six bodies ($\kappa=5$): reflex ($r$),
  synergy ($s$), and antagonist ($a$) coefficients of the shoulder (S),
  elbow (E), and biarticular (B) pools.}
  \label{tab:si-motif}
  \begin{tabular}{l ccc ccc ccc}
    \toprule
    Body & $r_{\mathrm S}$ & $r_{\mathrm E}$ & $r_{\mathrm B}$ & $s_{\mathrm S}$ & $s_{\mathrm E}$ & $s_{\mathrm B}$ & $a_{\mathrm S}$ & $a_{\mathrm E}$ & $a_{\mathrm B}$\\
    \midrule
    \multicolumn{10}{l}{\ding{73}\ $p_{\mathrm b}=0.0046$, $\lambda_{\max}=0.2$}\\
    SR$_0$, MR$_0$ & 0.93 & 0.98 & 0.96 & 0.38 & 0.22 & 0.29 & 0.01 & 0.03 & 0.02\\
    SR$_1$, MR$_0$ & 0.90 & 0.97 & 0.87 & 0.39 & 0.18 & 0.19 & 0.00 & 0.11 & 0.16\\
    SR$_2$, MR$_0$ & 0.95 & 0.88 & 0.76 & 0.13 & 0.00 & 0.16 & 0.09 & 0.26 & 0.23\\
    SR$_0$, MR$_1$ & 0.94 & 0.92 & 0.98 & 0.36 & 0.33 & 0.19 & 0.01 & 0.02 & 0.05\\
    SR$_1$, MR$_1$ & 0.75 & 0.95 & 0.87 & 0.48 & 0.36 & 0.23 & 0.00 & 0.00 & 0.08\\
    SR$_2$, MR$_1$ & 0.94 & 0.90 & 0.77 & 0.18 & 0.00 & 0.25 & 0.11 & 0.23 & 0.16\\
    \midrule
    \multicolumn{10}{l}{$\triangle$\ $p_{\mathrm b}=0.0129$, $\lambda_{\max}=0.5$}\\
    SR$_0$, MR$_0$ & 0.83 & 1.00 & 1.00 & 0.42 & 0.11 & 0.01 & 0.00 & 0.01 & 0.04\\
    SR$_1$, MR$_0$ & 0.47 & 1.00 & 0.55 & 0.69 & 0.30 & 0.01 & 0.00 & 0.00 & 0.32\\
    SR$_2$, MR$_0$ & 1.00 & 1.00 & 0.74 & 0.02 & 0.00 & 0.00 & 0.01 & 0.10 & 0.33\\
    SR$_0$, MR$_1$ & 0.82 & 1.00 & 1.00 & 0.42 & 0.12 & 0.01 & 0.00 & 0.01 & 0.04\\
    SR$_1$, MR$_1$ & 1.00 & 1.00 & 1.00 & 0.17 & 0.01 & 0.19 & 0.01 & 0.01 & 0.01\\
    SR$_2$, MR$_1$ & 1.00 & 1.00 & 0.75 & 0.02 & 0.00 & 0.00 & 0.01 & 0.10 & 0.33\\
    \midrule
    \multicolumn{10}{l}{$\Diamond$\ $p_{\mathrm b}=0.0028$, $\lambda_{\max}=0.9$}\\
    SR$_0$, MR$_0$ & 0.53 & 0.50 & 0.87 & 0.59 & 0.14 & 0.43 & 0.00 & 0.29 & 0.01\\
    SR$_1$, MR$_0$ & 0.72 & 1.00 & 0.00 & 0.50 & 0.28 & 0.00 & 0.00 & 0.00 & 0.51\\
    SR$_2$, MR$_0$ & 0.93 & 0.78 & 0.66 & 0.15 & 0.01 & 0.17 & 0.09 & 0.28 & 0.26\\
    SR$_0$, MR$_1$ & 0.49 & 0.49 & 0.89 & 0.61 & 0.19 & 0.40 & 0.01 & 0.30 & 0.01\\
    SR$_1$, MR$_1$ & 0.61 & 1.00 & 0.00 & 0.50 & 0.19 & 0.00 & 0.00 & 0.00 & 0.45\\
    SR$_2$, MR$_1$ & 0.93 & 0.76 & 0.69 & 0.17 & 0.01 & 0.18 & 0.09 & 0.29 & 0.25\\
    \bottomrule
  \end{tabular}
\end{table}

\begin{figure}[!ht]
  \centering
  \includegraphics[width=\linewidth]{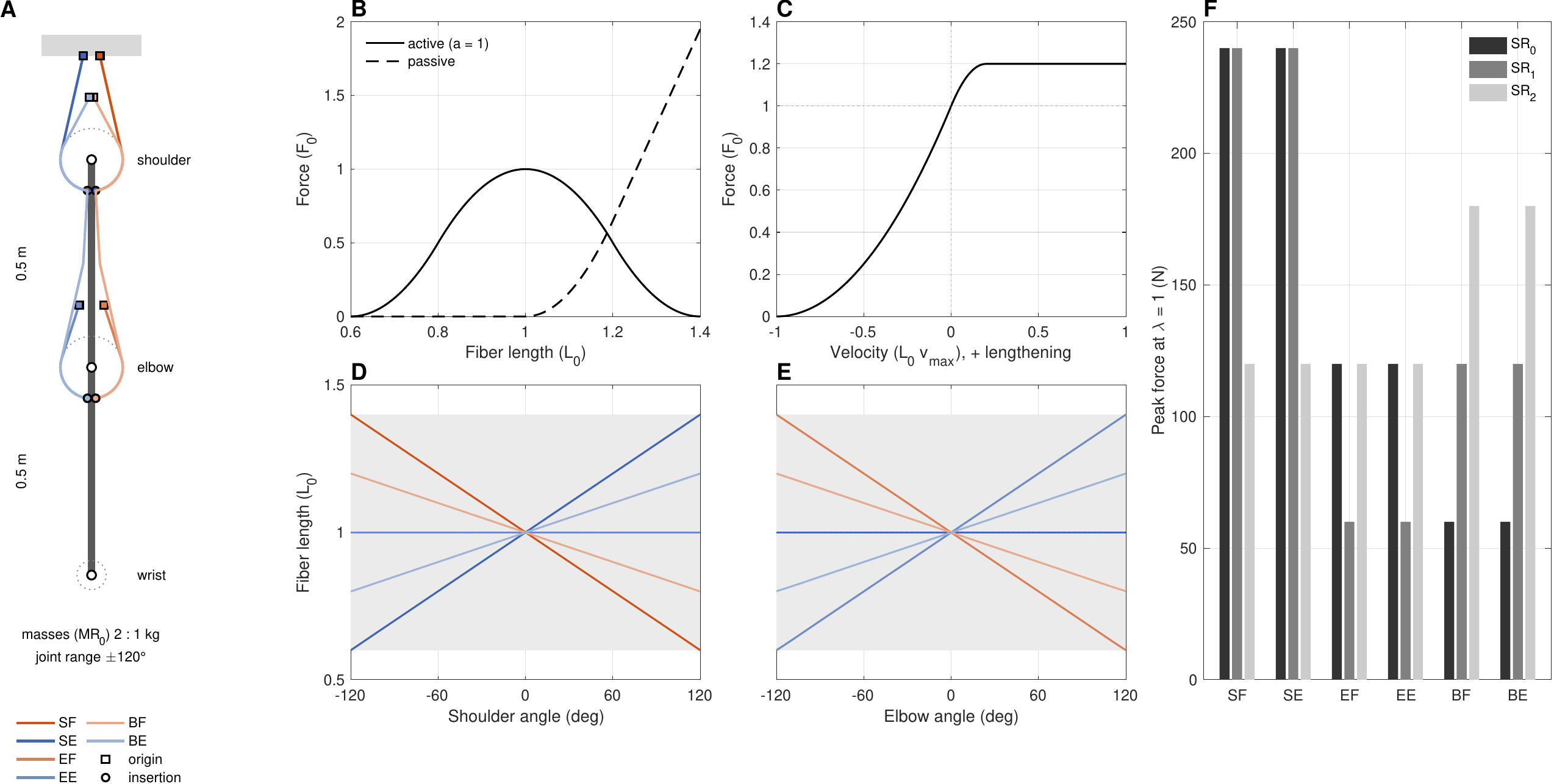}
  \caption{\textbf{Musculoskeletal plant}, computed from the MuJoCo model.
  (\textbf{A}) Planar two-segment arm at rest ($q=0$), shown hanging from the
  shoulder, with the paths of the six muscles from origin (squares) to
  insertion (circles); muscles wrap cylinders (dotted, radius 7.5~cm) at the
  shoulder and elbow, which gives every muscle a constant 7.5~cm moment arm
  about each joint it crosses. Segments are 0.5~m, with masses 2:1~kg
  (MR$_0$) and joint range $\pm120^\circ$; panel A is rotated by $90^\circ$
  from the model frame of Table~\ref{tab:si-geometry}, which lists the
  attachment and wrapping geometry of every muscle. (\textbf{B}) Active (activation
  $a=1$) and passive force--length relations and (\textbf{C}) force--velocity
  relation, in units of peak isometric force $F_0$. (\textbf{D},\textbf{E})
  Fiber length versus shoulder and elbow angle; shading marks the operating
  range $[0.6,1.4]\,\ell_0$, and at rest every muscle is at its optimal length
  $\ell_0$. (\textbf{F}) Peak isometric force of each muscle at full
  strength ($\lambda=1$) for the three strength ratios,
  $F_0=\lambda\,\rho_i\,F_{\mathrm{ref}}$ with $F_{\mathrm{ref}}=240$~N.}
  \label{fig:si-plant}
\end{figure}

\begin{table}[!ht]
  \centering
  \footnotesize
  \caption{\textbf{Muscle geometry.} Coordinates (m) at rest in the model
  frame (shoulder joint at the origin, arm along $+x$, flexor side $+y$);
  each point moves with the body it is fixed to. Paths wrap cylinders of
  radius 0.075~m centered on the shoulder $(0,0)$ and elbow $(0.5,0)$
  joints, on the side indicated.}
  \label{tab:si-geometry}
  \begin{tabular}{l ll l ll l}
    \toprule
    Muscle & \multicolumn{2}{l}{Origin} & Via points (upper arm) &
      \multicolumn{2}{l}{Insertion} & Wraps \\
    \midrule
    SF & base      & $(-0.25,\ 0.02)$  & ---                                & upper arm & $(0.075,-0.01)$ & shoulder, $+y$ \\
    SE & base      & $(-0.25,-0.02)$   & ---                                & upper arm & $(0.075,\ 0.01)$ & shoulder, $-y$ \\
    EF & upper arm & $(0.35,\ 0.03)$   & ---                                & forearm   & $(0.575,\ 0.01)$ & elbow, $+y$ \\
    EE & upper arm & $(0.35,-0.03)$    & ---                                & forearm   & $(0.575,-0.01)$ & elbow, $-y$ \\
    BF & base      & $(-0.15,\ 0.005)$ & $(0.075,\ 0.01)$, $(0.25,\ 0.02)$ & forearm   & $(0.575,\ 0.01)$ & shoulder, elbow, $+y$ \\
    BE & base      & $(-0.15,-0.005)$  & $(0.075,-0.01)$, $(0.25,-0.02)$   & forearm   & $(0.575,-0.01)$ & shoulder, elbow, $-y$ \\
    \bottomrule
  \end{tabular}
\end{table}

\begin{figure}[!ht]
  \centering
  \includegraphics[width=0.5\linewidth]{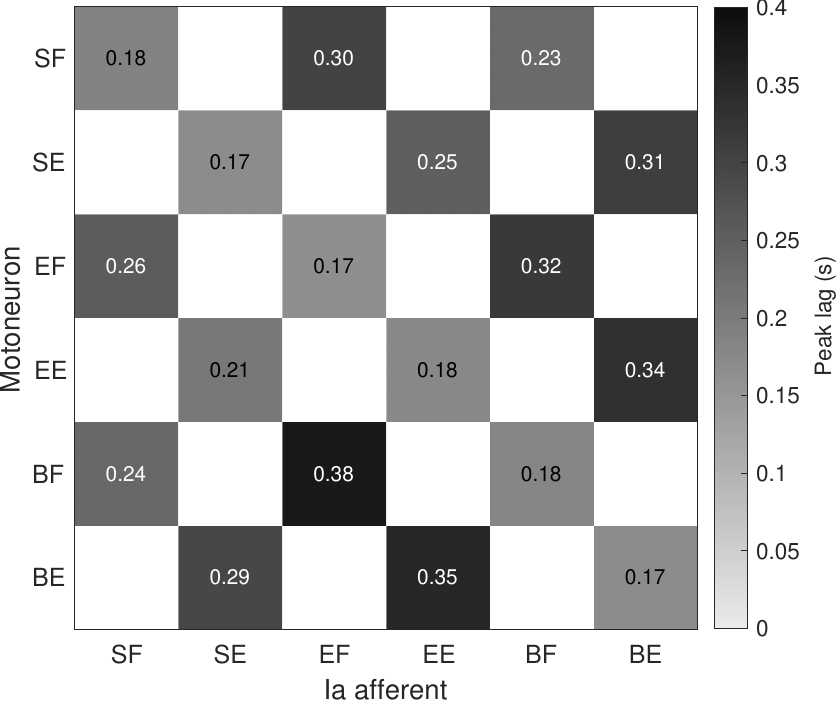}
  \caption{\textbf{Delay from presynaptic afferent to weight change.} For each
  homonymous and synergist connection of the canonical simulation (rows,
  motoneurons; columns, Ia afferents), the lag at which the Ia afferent
  activity $x_{\mathrm{Ia},j}$ best predicts the weight change
  $\dot w_{ij}$: the peak of their cross-correlation over the full
  simulation, searched up to 0.5~s with the afferent leading. Homonymous
  connections peak at short, uniform lags (0.17--0.19~s); synergist
  connections peak later (0.21--0.38~s), the longest involving elbow and
  biarticular muscles. Peak correlations are modest ($|r|=0.13$--0.22).
  Antagonist connections, which decay, are not shown. The longest lag sets
  the 0.4~s history window of the Granger causality analysis
  (Fig.~\ref{fig:gc-partialR2}).}
  \label{fig:si-ia-lag}
\end{figure}